\documentclass[%
 aip,
 amsmath,amssymb,
 reprint,%
]{revtex4-1}

\usepackage{graphicx}% Include figure files
\usepackage{dcolumn}% Align table columns on decimal point

\usepackage[utf8]{inputenc}
\usepackage[T1]{fontenc}
\usepackage{braket}
\usepackage{amsmath, amssymb}
\usepackage{mathtools} % enable matrix alignment
\usepackage{upgreek}
\usepackage{xspace} % enable space between unit
\usepackage{bbold} % enable \mathbb{1}
\usepackage{overpic}
\usepackage{fvextra}  % extends fancyvrb
\usepackage{hyperref}
\usepackage{here}
\makeatletter
\def\@email#1#2{%
 \endgroup
 \patchcmd{\titleblock@produce}
  {\frontmatter@RRAPformat}
  {\frontmatter@RRAPformat{\produce@RRAP{*#1\href{mailto:#2}{#2}}}\frontmatter@RRAPformat}
  {}{}
}%
\makeatother

\begin{document}
\preprint{AIP/123-QED}

\newcommand{\us}{\ensuremath{\upmu\text{s}}\xspace}
\newcommand{\ns}{\ensuremath{\text{ns}}\xspace}
\newcommand{\rad}{\ensuremath{\text{rad}}\xspace}
\newcommand{\usinv}{\ensuremath{\upmu\text{s}^{-1}}\xspace}
\newcommand{\sinv}{\ensuremath{\text{s}^{-1}}\xspace}
\newcommand{\nsinv}{\ensuremath{\text{ns}^{-1}}\xspace}
\newcommand{\mT}{\ensuremath{\text{mT}}\xspace}
\newcommand{\mTinv}{\ensuremath{\text{mT}^{-1}}\xspace}

% length unit: Angstrom (math-safe)
\newcommand{\Ang}{\ensuremath{\text{\AA}}\xspace}

% for superoperator |a>>
\newcommand{\dket}[1]{\ensuremath{\left\lvert #1 \right\rangle\rangle}}
\newcommand{\dbra}[1]{\ensuremath{\left\langle\langle #1 \right\rvert}}
\newcommand{\dhat}[1]{\ensuremath{\hat{\hat{#1}}}}

% for atom
\newcommand{\Hyd}[1]{\ensuremath{{}^{#1}\text{H}}\xspace}
\newcommand{\Nit}[1]{\ensuremath{{}^{#1}\text{N}}\xspace}
\newcommand{\Trp}[1]{\ensuremath{\text{Trp}_{\text{#1}}}\xspace}
\newcommand{\RP}[1]{\ensuremath{\text{RP}_{\text{#1}}}\xspace}

\title{Introduction to modelling radical pair quantum spin dynamics with tensor networks}

\author{Kentaro Hino}
\altaffiliation[Electronic mail: ]{hino@theoc.kuchem.kyoto-u.ac.jp}
\affiliation{Department of Chemistry, Graduate School of Science, Kyoto University, Kitashirakawa Oiwake-cho, Sakyo, Kyoto, 606-8502, Japan}

\author{Damyan S. Frantzov}%
\affiliation{Department of Chemistry, Chemistry Research Laboratory, University of Oxford, Oxford, OX1 3TA, United Kingdom}

\author{Yuki Kurashige}
\affiliation{Department of Chemistry, Graduate School of Science, Kyoto University, Kitashirakawa Oiwake-cho, Sakyo, Kyoto, 606-8502, Japan}
\affiliation{CREST, JST, Honcho 4-1-8, Kawaguchi, Saitama, 332-0012, Japan}

\author{Lewis M. Antill}
\altaffiliation[Electronic mail: ]{lewis.antill@g.skku.edu}
\affiliation{Institute of Quantum Biophysics, Department of Biophysics, Sungkyunkwan University, Suwon, 16419, Republic of Korea}
%\email{lewis.antill@g.skku.edu}

\date{\today}% It is always \today, today,

\begin{abstract}
Radical pairs are short-lived, spin-correlated intermediates that underpin processes in chemistry, biology, and emerging quantum technologies. Their behaviour is governed by coupled electron-nuclear spin dynamics and is sensitive to weak magnetic fields, but full quantum treatments have been obstructed by the extreme computational cost of modelling many interacting spins. Here, this barrier is removed, demonstrating that open-system radical-pair dynamics can be resolved at nuclear-spin scales previously intractable, explicitly reaching regimes with tens of coupled nuclei and validated up to 60 spins. In biologically relevant flavin-tryptophan radical pairs, electron-transfer pathways and magnetic-field anisotropy are found to reshape spin evolution and, in turn, spin-selective reaction yields. The resulting directional responses expose a strong mechanistic link between the nuclear environment, magnetic geometry, and chemical outcome \textemdash \ a relationship central to hypotheses of avian magnetoreception and other magnetic-field effects in biology. This establishes a widely applicable simulation framework that removes a long-standing barrier in spin chemistry, quantum biology, and spin-based device science.
\end{abstract}

\maketitle

\section*{Introduction}
Spin relaxation in molecular systems governs key mechanistic pathways across quantum computing architectures, spintronic devices, and quantum-biophysical processes. In magnetic-resonance experiments, controlling relaxation pathways provides an additional lever for extracting electronic and nuclear spin information. Radical-pair relaxation dynamics, in particular, underpin excited-state reactivity in organic semiconductors \cite{dediuSpinRoutesOrganic2009, weissStronglyExchangecoupledTriplet2017},
define coherence lifetimes in molecular qubits \cite{divincenzoPhysicalImplementationQuantum2000, chatterjeeSemiconductorQubitsPractice2021},
and are implicated in the magnetic compass sense of migratory birds \cite{ritzMagneticCompassBirds2009,horeRadicalPairMechanismMagnetoreception2016}.

A radical pair typically comprises two unpaired electrons coupled to tens of surrounding nuclei and subjected to external magnetic fields of arbitrary magnitudes and orientations.
Existing theoretical frameworks span classical descriptions of the nuclear bath \cite{schultenSemiclassicalDescriptionElectron1978, manolopoulosImprovedSemiclassicalTheory2013, p.fayHowQuantumRadical2020, antillRadicalPyToolSpin2024}
%\cite{schulten1978wolynes, manolopoulos2013hore, p.fay2020j.hore, antill2024vatai}
to full quantum models treating both nuclear and electronic spins \cite{faySpinRelaxationRadical2021}.
Three barriers set the practical limits of simulation:
(i)\;the nuclei occupy a room-temperature mixed state, requiring ensemble averaging over many initial conditions;
(ii)\;the Hilbert and Liouville spaces expand exponentially with increasing nuclei;
(iii)\;spin relaxation unfolds on 1 ns to 1 \us time scales, necessitating numerically stable long-time propagation.

Challenge (i) is often handled via Monte-Carlo sampling over initial nuclear configurations \cite{lewisEfficientQuantumMechanical2016}.
Stochastic Schrödinger approaches \cite{faySpinRelaxationRadical2021} retain the full quantum structure but still confront the exponential scaling in (ii).
Classical-vector approximations alleviate this burden \cite{schultenSemiclassicalDescriptionElectron1978, manolopoulosImprovedSemiclassicalTheory2013} and reliably capture magnetic-field effects at high fields or large bath sizes, though a fully quantum treatment becomes essential when non-Markovian hyperfine-mediated relaxation is significant\cite{p.fayHowQuantumRadical2020}.

With respect to (iii), ultrafast vibrational modes (fs-ps) are commonly coarse-grained into Markovian decay channels, as in the Haberkorn model \cite{haberkornDensityMatrixDescription1976} and Bloch-Redfield-Wangsness theory \cite{redfieldTheoryRelaxationProcesses1965, wangsnessDynamicalTheoryNuclear1953}.
Working in Liouville space removes the need for ensemble averaging and enables general relaxation superoperators or Lindblad jumps, whereas Hilbert-space stochastic schemes incorporate similar physics via quantum-jump trajectories \cite{keensMonteCarloWavefunctionApproach2020}.

Tensor networks offer a natural route past exponential complexity in many-body quantum dynamics. In particular, matrix-product-state (MPS) methods based on the density-matrix renormalisation group
\cite{whiteDensityMatrixFormulation1992c}
and time-evolving block decimation (TEBD) \cite{vidalEfficientClassicalSimulation2003b, whiteRealTimeEvolutionUsing2004b}
have become standard and have reached isotropic central-spin systems with up to 999 bath spins
\cite{stanekDynamicsDecoherenceCentral2013}.

More recently, projector-splitting integrators have been derived from the time-dependent variational principle (TDVP)
\cite{haegemanTimeDependentVariationalPrinciple2011b, haegemanUnifyingTimeEvolution2016b, lubichDynamicalApproximationHierarchical2013a}
enable real-time evolution for Hamiltonians with long-range couplings, beyond the nearest-neighbour scope of TEBD.

Exact open-system methods such as the hierarchical equations of motion \cite{tanimuraNumericallyExactApproach2020} and tensor-network simulations of the spin-boson model via TDVP \cite{schroderSimulatingOpenQuantum2016} further illustrate the reach of these approaches; however, radical-pair applications have remained limited due to the required dual expertise in open-quantum tensor methods and spin chemistry.

\begin{figure}%[H]%[tbp]
    \centering
    \includegraphics[width=0.98\linewidth]{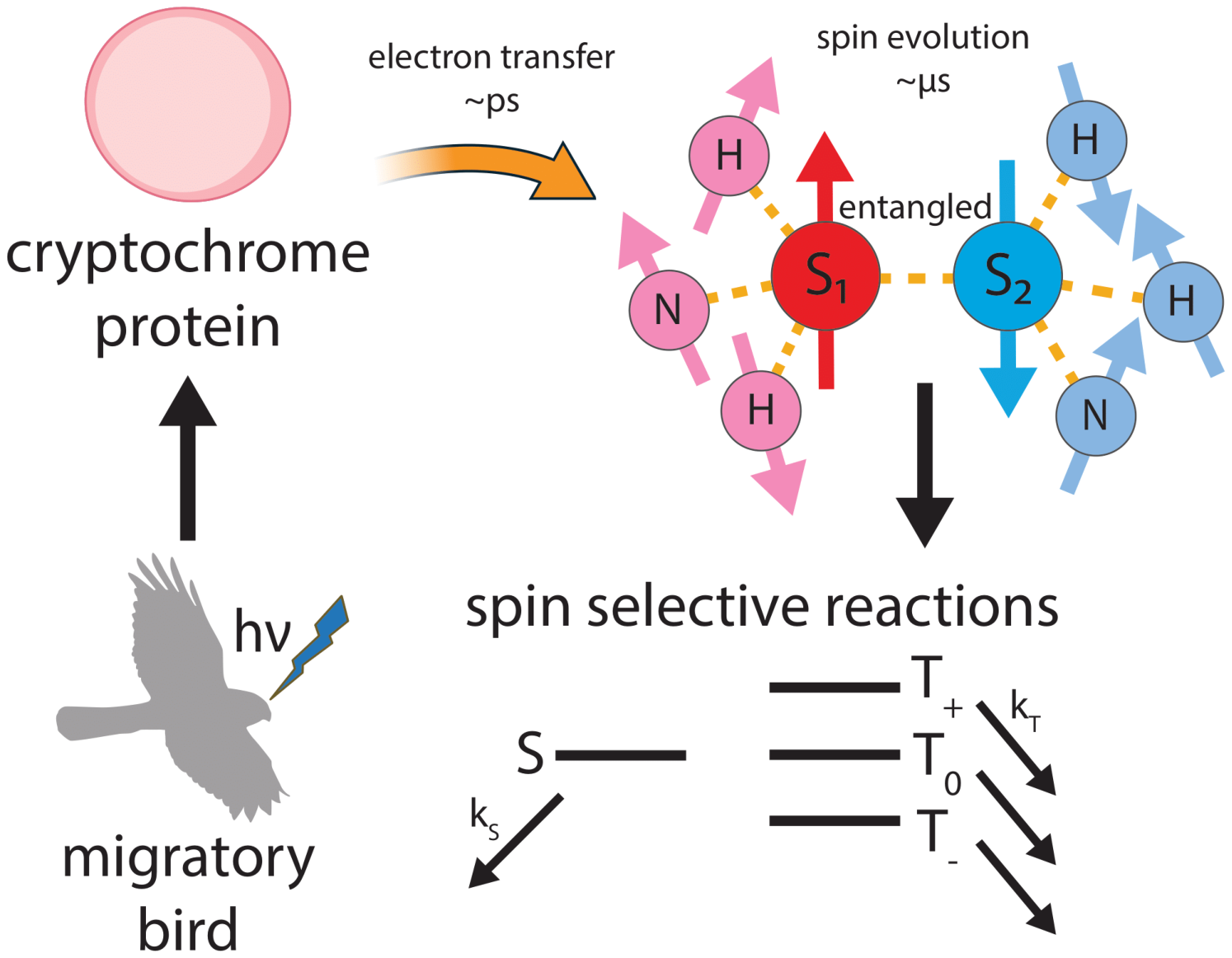}
    \caption{\label{fig:qbirds}Radical-pair model and workflow. Schematic of a typical radical-pair system. Spin evolution is computed with tensor-network simulators (MPS, vMPDO, LPMPS), and spin-selective reaction yields are obtained from a kinetic scheme driven by cumulative (time-integrated) state populations produced by the spin dynamics.}
\end{figure}

This work, therefore, pursues three aims: (i) to introduce and benchmark modern tensor-network techniques within spin chemistry and delineate the scale of problems now tractable; (ii) to articulate the radical-pair problem to the broader physics community and clarify its intrinsic challenges; and (iii) to demonstrate a biologically consequential quantum-spin phenomenon \textemdash \ avian magnetoreception \textemdash \ as an application context (see Fig. \ref{fig:qbirds}).

\section{Three tensor network methods for radical pair mixed states}
We first recall the structure of the radical-pair Hilbert space before introducing the three tensor-network representations used in this work. The full Hilbert space is spanned by (i) the electronic spin basis $\ket{\sigma_{\mathrm{el}}} \in \{\ket{T_+}, \ket{T_0}, \ket{S}, \ket{T_-}\}$, and (ii) the nuclear spin basis $\ket{\sigma_{\mathrm{nuc}}^{(i,j)}} \in \left\{\ket{I^{(i,j)}; I^{(i,j)}}, \ket{I^{(i,j)}; I^{(i,j)}-1}, \cdots, \ket{I^{(i,j)}; -I^{(i,j)}}\right\}$, where
$I^{(i,j)}$ is the spin quantum number of the $j$-th nucleus coupled to electron $i$. The two-electron basis $\{\ket{\uparrow\uparrow}, \ket{\uparrow\downarrow}, \ket{\downarrow\uparrow}, \ket{\downarrow\downarrow}\}$ is rotated into the singlet-triplet basis by
\begin{equation}
\begin{aligned}
\ket{T_+} &:= \ket{\uparrow\uparrow}, \\
\ket{T_0} &:= \frac{1}{\sqrt{2}} \left( \ket{\uparrow\downarrow} + \ket{\downarrow\uparrow} \right), \\
\ket{T_-} &:= \ket{\downarrow\downarrow} \\
\ket{S} &:= \frac{1}{\sqrt{2}} \left( \ket{\uparrow\downarrow} - \ket{\downarrow\uparrow} \right)
.
\end{aligned}
\end{equation}
The total Hamiltonian is written as
\begin{equation}
\begin{aligned}
\hat{H}_{\mathrm{total}}
&= \hat{H}_{\mathrm{Z}} + \hat{H}_{\mathrm{H}} + \hat{H}_{\mathrm{J}} + \hat{H}_{\mathrm{D}} + \hat{H}_{\mathrm{K}}
\end{aligned}
\end{equation}
with the constituent terms
\begin{equation}
\begin{aligned}
  \hat{H}_{\mathrm{Z}} &= -\mathbf{B}^\top \cdot
  \sum_{i=1}^{2}
  \left(
  \gamma^{(\mathrm{e})} \hat{\mathbf{S}}_i
  + \sum_{j=1}^{N_i} \gamma_{i,j}^{(\mathrm{n})} \hat{\mathbf{I}}_{i,j}
  \right) \\
  \hat{H}_{\mathrm{H}} &= |\gamma^{(\mathrm{e})}|
  \sum_{i=1}^{2} \sum_{j=1}^{N_i} \hat{\mathbf{S}}_i^\top \cdot \mathbf{A}_{i,j} \cdot \hat{\mathbf{I}}_{i,j} \\
  \hat{H}_{\mathrm{J}} &= -J|\gamma^{(\mathrm{e})}| \left(
    2 \hat{\mathbf{S}}_1^\top \cdot \hat{\mathbf{S}}_2 - \frac{1}{2}\hat{\mathbb{1}}
  \right) \\
  \hat{H}_{\mathrm{D}}
  &=|\gamma^{(\mathrm{e})}| \hat{\mathbf{S}}_1^\top \cdot \mathbf{D} \cdot \hat{\mathbf{S}}_2 \\
  \hat{H}_\mathrm{K} &= - \frac{i}{2}\left(k_S \hat{P}_S + k_T \hat{P}_T\right)
\end{aligned}
\end{equation}
where $\hat{H}_{\mathrm{Z}}$, $\hat{H}_{\mathrm{H}}$, $\hat{H}_{\mathrm{J}}$, $\hat{H}_{\mathrm{D}}$,
$\hat{H}_{\mathrm{K}}$ represent the Zeeman, hyperfine, exchange, dipolar, and Haberkorn terms, respectively.
Here $i\in \{1,2\}$ indexes electrons and $j\in\left\{1,2,\cdots,N_i\right\}$ indexes nuclei.
The spin operators are $\hat{\mathbf{S}}_i = \left[ \hat{\mathrm{S}}^{i}_x, \hat{\mathrm{S}}^{i}_y, \hat{\mathrm{S}}^{i}_z \right]^\top$  and $\hat{\mathbf{I}}_{i,j} = \left[ \hat{\mathrm{I}}^{i,j}_x, \hat{\mathrm{I}}^{i,j}_y, \hat{\mathrm{I}}^{i,j}_z \right]^\top$;
$\hat{P}_S = \frac{1}{4}\hat{\mathbb{1}}_4 - \hat{\mathbf{S}}_1^\top \cdot \hat{\mathbf{S}}_2$ and $\hat{P}_T = \hat{\mathbb{1}}_4 - \hat{P}_S$ are the singlet/triplet projection operators.
The gyromagnetic ratios are
$\gamma^{(\mathrm{e})}$ (electrons)
and $\gamma_{i,j}^{(\mathrm{n})}$ (nuclei).
We set $\hbar=1$ and work in energy units scaled by $|\gamma^{(\mathrm{e})}|\hbar$.
The parameters are: magnetic field
$\mathbf{B} \in \mathbb{R}^3$, hyperfine tensors
$\mathbf{A}_{i,j} \in \mathbb{R}^{3\times 3}$, exchange coupling constant $J\in \mathbb{R}$, dipolar tensor $\mathbf{D} \in \mathbb{R}^{3\times 3}$, and spin-selective reaction rates $k_S$ and $k_T$.

Observable radical-pair outputs are typically electronic-state populations, written as
\begin{equation}
  \label{eq:pop-observable}
  \begin{aligned}
    &\langle P_X(t) \rangle \\
    &= \frac{1}{Z}\mathrm{Tr}\left(\hat{P}_X\hat{\rho}(t)\right)\\
    &= \frac{1}{Z} \mathrm{Tr} \left(\hat{P}_X U(t) \ket{\Psi_{\mathrm{ele}}(0)} \bra{\Psi_{\mathrm{ele}}(0)}U(t)^\dagger \right) \\
    &= \int \mathrm{d}\boldsymbol{\Omega} p(\boldsymbol{\Omega}) \Braket{\Psi_{\mathrm{ele}}(0) \boldsymbol{\Omega} | \hat{U}(t)^\dagger \hat{P}_X \hat{U}(t) | \Psi_{\mathrm{ele}}(0) \boldsymbol{\Omega}},
  \end{aligned}
\end{equation}
In stochastic wavefunction or classical-vector approaches, the trace is replaced by Monte-Carlo insertion of the identity into
$\ket{\Psi_{\mathrm{ele}}(0)}\hat{\mathbb{1}}\bra{\Psi_{\mathrm{ele}}(0)}$, sampling initial nuclear configurations $\boldsymbol{\Omega}$.

We now describe the three one-dimensional tensor network ans{\"a}tze used.

(1) Stochastic MPS (Hilbert-space ensemble):
\begin{equation}
    \begin{aligned}
  \sum_{\sigma_1, \sigma_2, \cdots, \sigma_N} \sum_{\alpha_1, \cdots, \alpha_{N-1}} A\substack{\sigma_1\\\alpha_1} A\substack{\sigma_2\\\alpha_1\alpha_2} \cdots A\substack{\sigma_N\\\alpha_{N-1}}
  \ket{\sigma_1, \sigma_2, \cdots, \sigma_N},
  \end{aligned}
\end{equation}
where the local basis sites $\ket{\sigma_k}$ enumerate electronic and nuclear degrees of freedom, and the expectation values are taken over Monte-Carlo trajectories.

(2) Vectorised MPDO (Liouville-space tensor-train):
\begin{equation}
\begin{gathered}
  \sum_{\rho_1, \rho_2, \cdots, \rho_N}
  \sum_{\gamma_1, \cdots, \gamma_{N-1}}
  C\substack{\rho_1\\\gamma_1}
  C\substack{\rho_2\\\gamma_1\gamma_2}
  \cdots
  C\substack{\rho_N\\\gamma_{N-1}}
  \dket{\rho_1, \rho_2, \cdots, \rho_N},
\end{gathered}
\end{equation}
with the local Liouville basis $\dket{\rho_k} := \mathrm{vec}(\ket{\sigma_k^\prime}\bra{\sigma_k}) \in \mathbb{C}^{d_k^2}$. This low-rank representation can violate complete positivity or trace preservation if the bond dimension $\chi$ is insufficient.

(3) Locally purified MPS (deterministic MPDO via ancillas):
\begin{equation}
   \hat{\rho}_{\mathrm{phys}}(t) = \mathrm{Tr}_{\{s_1, s_2, \cdots, s_N\}} \left\{\ket{\Psi(t)}\bra{\Psi(t)}\right\},
\end{equation}
where $\ket{\Psi(t)}$ is a matrix product state spanned by physical sites $\{\ket{\sigma_k}\}$ and ancilla sites $\{\ket{s_k}\}$
\begin{equation}
  \sum_{\{\boldsymbol{\sigma}\}, \{\boldsymbol{s}\}, \{\boldsymbol{\alpha}\}}
    A\substack{\sigma_1\\\alpha_1}
    A\substack{s_1\\\alpha_1\alpha_2}
    A\substack{\sigma_2\\\alpha_2\alpha_3}
    \cdots
    A\substack{s_N\\\alpha_{2N-1}}
  \ket{\sigma_1, s_1, \sigma_2, \cdots, s_N}.
\end{equation}
This approach enforces positivity by construction through partial tracing over ancilla sites. We denote the bond dimensions of MPS, vMPDO and LPMPS by $m$, $\chi$, and $r$, respectively.
Diagrammatic representations are shown in Fig.~\ref{fig:mps-mpo}, with algorithmic specifics provided in Appendix \ref{sec:methods}.

\begin{figure}
  \centering
  \begin{overpic}[width=.98\linewidth]{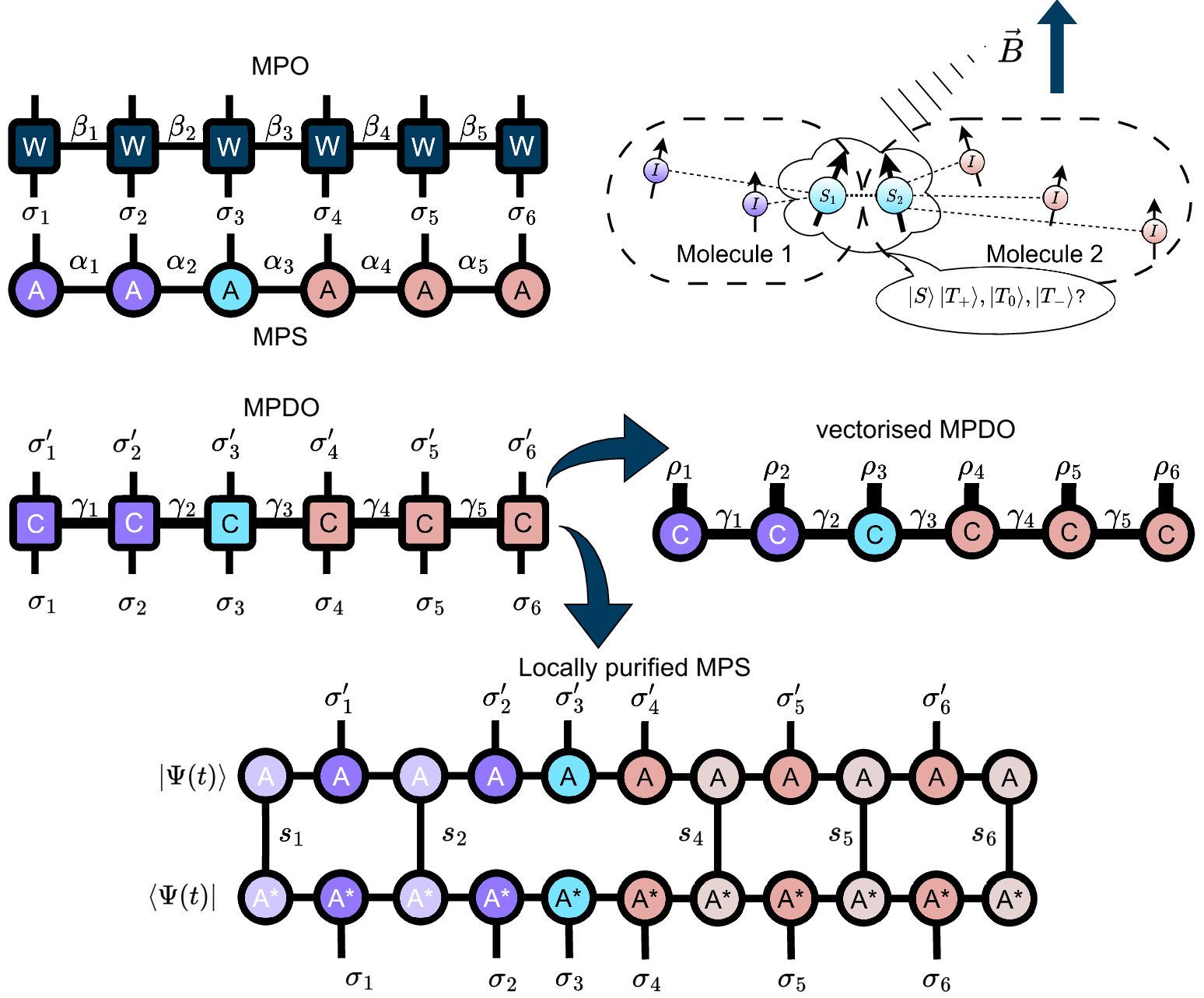}
    \put(1,77){\textbf{a}} % x=5 %, y=4 % from bottom-left
  \end{overpic}
  \newline
  \newline
  \begin{overpic}[width=.98\linewidth]{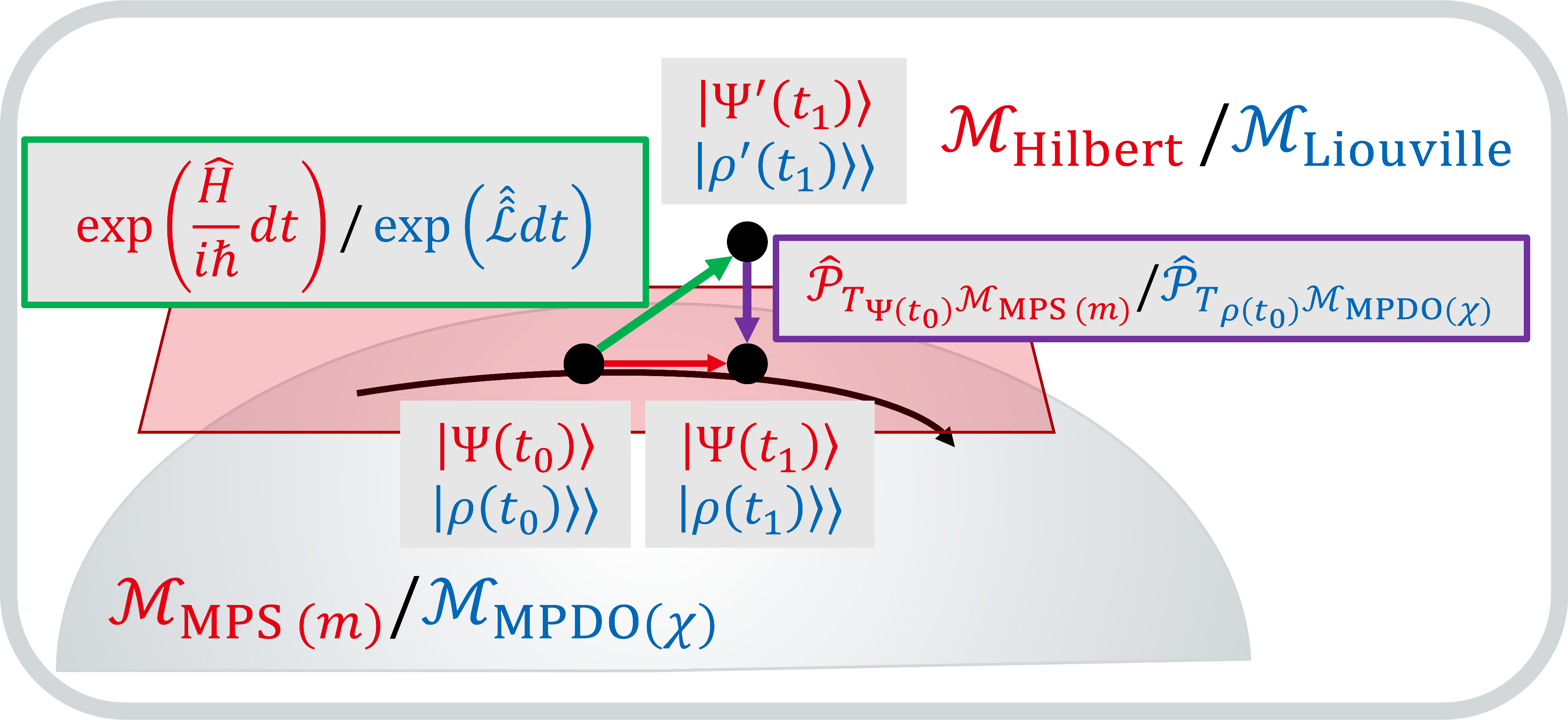}
    \put(1,48){\textbf{b}} % x=5 %, y=4 % from bottom-left
    \hfill
  \end{overpic}
  \caption{Tensor-network representations and time evolution.
    \textbf{(a)} Schematic tensor-network layout for a radical-pair system with five nuclear spins. Purple nodes denote nuclear spins on molecule $i=1$, light-blue nodes denote the two electronic-spin sites, and orange nodes denote nuclear spins on molecule $i=2$.
    \textbf{(b)} Illustration of time evolution on the MPS/MPDO manifold via tangent-space projection under the time-dependent variational principle.
  }
  \label{fig:mps-mpo}
\end{figure}

\section{Convergence behaviour for an 18-nuclear-spin system}
\label{subsec:convergence}
\subsection{Spin model specification}
We benchmark both stochastic (full wavefunction and MPS) and deterministic (vMPDO and LPMPS) approaches for a radical pair consisting of 18 nuclear spins in a flavin anion ($i=1$) and a tryptophan cation ($i=2$), for which the Hilbert space dimension is $5,308,416$.%, which is close to the maximum size that can be simulated on a typical single computational node by exploiting the sparsity of the Hamiltonian and Krylov subspace methods.

Isotropic hyperfine tensors $\mathbf{A}_{i,j} = a_{i,j}\hat{\mathbb{1}}_3$ (listed in Table~\ref{tab:hyperfine}) are used.
The exchange coupling is set to $J  = 0.224\;\mT$ and the dipolar interaction is set to $\mathbf{D}=-0.38\times\mathrm{diag}\left[-\tfrac{2}{3}, -\tfrac{2}{3}, \tfrac{4}{3}\right]\;\mT$ \cite{p.fayHowQuantumRadical2020,schweigerPrinciplesPulseElectron2001}.
A Zeeman field is applied along the $z$-axis at strengths
$0.05\;\mT$, $0.50\;\mT$, and $5.00\;\mT$.

The lowest field ($0.05\;\mT$) is comparable to the geomagnetic field and slightly splits $\ket{T_{+}}$ and $\ket{T_{-}}$; the intermediate field ($0.50\;\mT$) approaches $2J$ so that the Zeeman gap compensates for the exchange splitting between $\ket{S}$ and $\ket{T_{+}}$; and the high field ($5.00\;\mT$) isolates $\ket{T_+}$ and $\ket{T_-}$ from $\ket{S}$ and $\ket{T_0}$.
Spin-selective decay rates in the Haberkorn model are set to $k_S = k_T = 1\;\usinv$.
Initial nuclear configurations for the full-wavefunction approach are generated using SU(Z) sampling \cite{faySpinRelaxationRadical2021,nemotoGeneralizedCoherentStates2000}, whereas for stochastic MPS, we employ spin-coherent-state sampling
\cite{radcliffePropertiesCoherentSpin1971}.
For all stochastic simulations, we find $K=4096$ trajectories to be sufficient for convergence; the dependence on $K$ is analysed in Appendix \ref{subsec:Kdeps}.

\subsection{Importance of a fully quantum-mechanical treatment}
To assess the intrinsic difficulty of this system, Fig.~\ref{fig:mconv}a compares classical-vector approaches \textemdash \ Schulten-Wolynes (SW) theory \cite{schultenSemiclassicalDescriptionElectron1978} and semiclassical (SC) theory \cite{manolopoulosImprovedSemiclassicalTheory2013,p.fayHowQuantumRadical2020} \textemdash \ against fully quantum simulations (stochastic MPS with bond dimension $m=16$ and the full wavefunction reference).
SC equations of motion and observables are given in Appendix \ref{subsec:eom-sc}.
For the classical methods, we averaged over $10^{7}$ trajectories to ensure convergence.
At low field ($|B_z|=0.05\;\mT$), SW performs poorly, whereas SC reproduces the qualitative trend but misses short-time oscillations. By contrast, MPS with $m=16$ accurately captures the fluctuations and exhibits smaller deviations than SC even at long times. Thus, for this regime, MPS at modest $m$ outperforms SC in both accuracy and computational efficiency.
\subsection{Dependence on tensor-network bond dimensions}
We next examine the convergence with respect to the bond dimensions of MPS ($m$), vMPDO ($\chi$), and LPMPS ($r$).
Fig.~\ref{fig:mconv}b shows the bond-dimension dependence at $|B_z| = 0.05\;\mT$. Achieving 0.5\% absolute accuracy in the singlet population at $t=200$ ns requires $m=128$ with $K=4096$ samples for MPS and $\chi=1024$ for vMPDO.
For all tensor network methods, small bond dimensions induce non-physical oscillations, which are particularly pronounced in vMPDO.
Deterministic approaches (vMPDO and LPMPS) demand larger bond dimensions than stochastic MPS at comparable accuracy.

All methods agree well at short times ($t \lesssim 100\;$ns), as the wavefunction and density matrix initially possess an effectively rank-1 structure (i.e., $m=\chi=1$ for MPS/vMPDO and $r=\max{(2I_{i,j}+1)}=3$ for LPMPS).
As evolution proceeds, entanglement grows, and the quantum state spreads over larger Hilbert/Liouville-space manifolds, eventually exceeding the representational capacity of low-rank tensor networks.
For reference, the full-wavefunction corresponds to the limiting bond dimension $m=1536$ for this problem, enabling a direct quantification of the accuracy trade-off.

\subsection{Magnetic field dependence of convergence}
\begin{figure*}
  \centering
  \begin{overpic}[width=.85\linewidth]{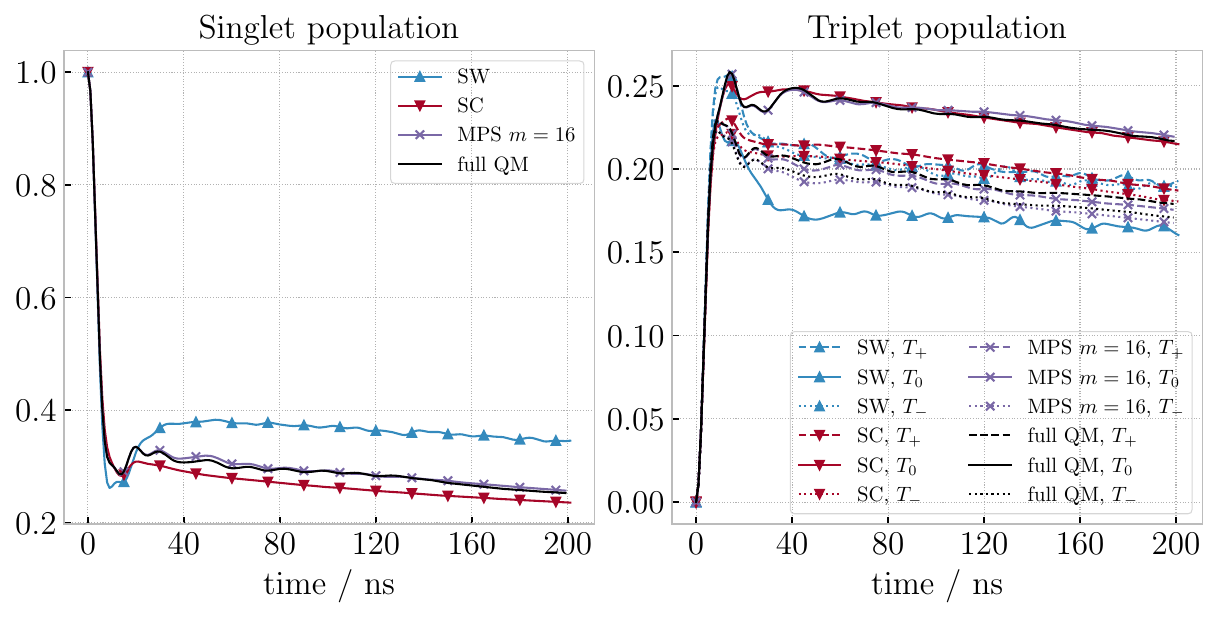}
    \put(1,50){\textbf{a}} % x=5 %, y=4 % from bottom-left
  \end{overpic}
  \begin{overpic}[width=.85\linewidth]{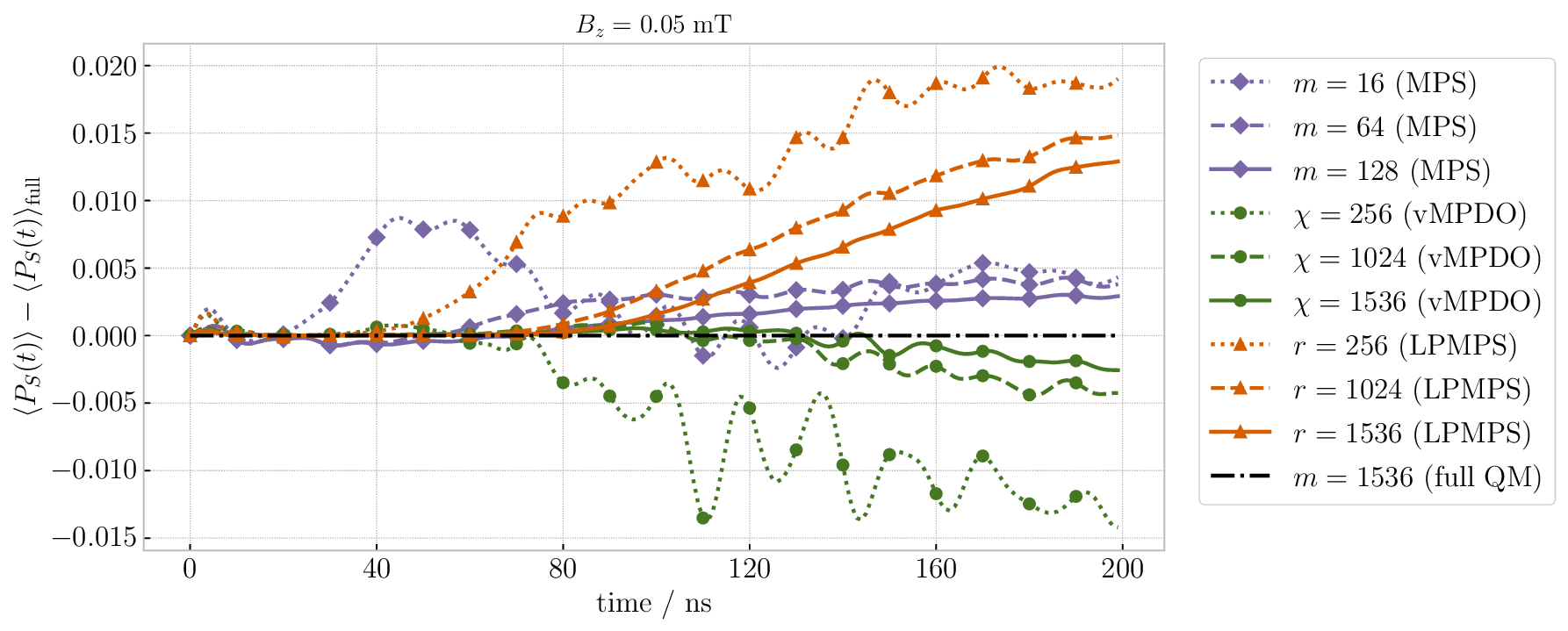}
    \put(1,40){\textbf{b}} % x=5 %, y=4 % from bottom-left
  \end{overpic}
  \begin{overpic}[width=.85\linewidth]{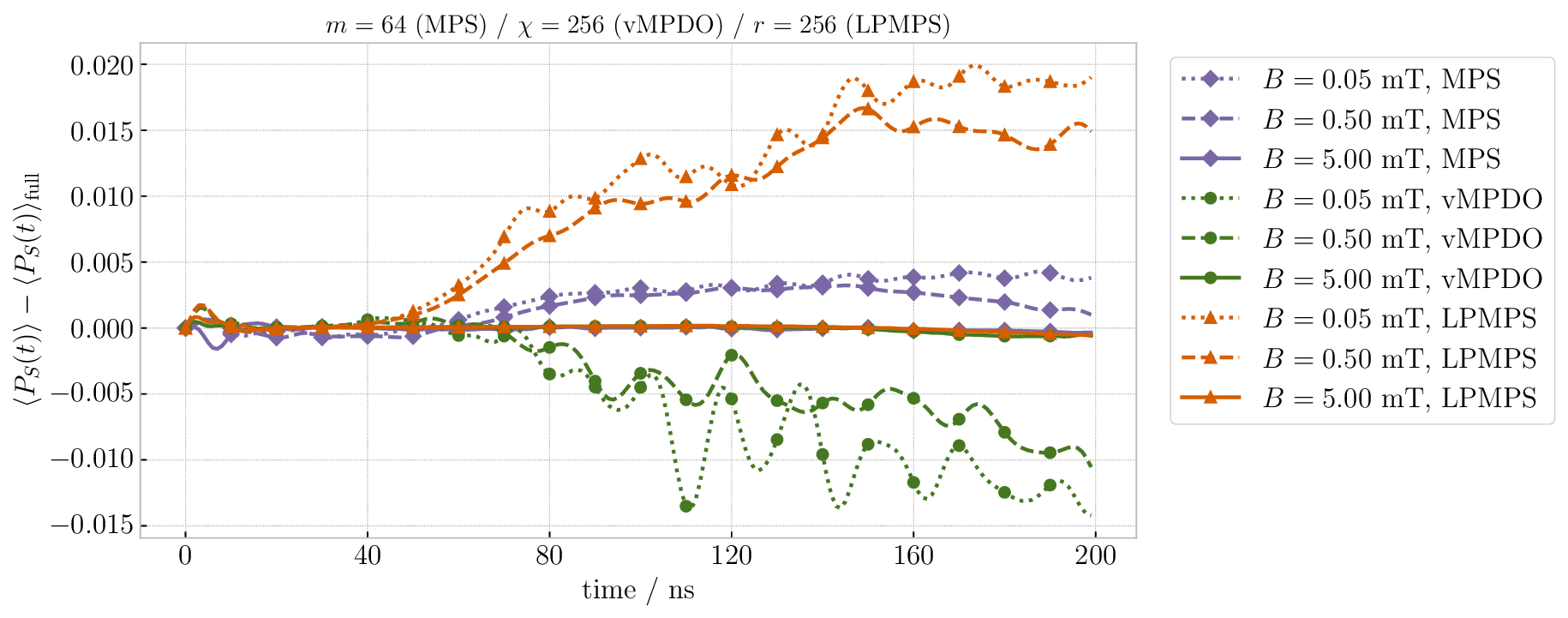}
    \put(1,40){\textbf{c}} % x=5 %, y=4 % from bottom-left
  \end{overpic}
  \caption{Convergence and classical-quantum comparison for an 18-nuclear-spin radical pair. \textbf{(a)} Population dynamics from classical-vector approaches (Schulten-Wolynes, SW; semiclassical, SC) compared against tensor-network MPS and the full wavefunction reference.
  \textbf{(b)} Convergence with bond dimension for $m\in\{16, 64, 128\}$ (stochastic MPS with $K=4096$ trajectories) and $\chi, r \in \{256, 1024, 1536\}$ (vMPDO/LPMPS). Each trace shows the trajectory-averaged diagonal of the reduced density matrix.
  \textbf{(c)} Dependence of convergence on magnetic-field strength.
  }
  \label{fig:mconv}
\end{figure*}

Fig~\ref{fig:mconv}c shows the convergence behaviour at fixed bond dimension across different field strengths. Higher magnetic fields yield faster and more reliable convergence. This is consistent with the known result that classical treatments of the nuclear bath better reproduce magnetic-field responses in the high-field limit \cite{manolopoulosImprovedSemiclassicalTheory2013}.

Physically, increasing the external field suppresses nuclear-electron entanglement because hyperfine, exchange, and dipolar couplings become small relative to the dominant single-body Zeeman term. In the limiting case where exchange and dipolar interactions vanish, the two-electron problem factorises into two independent central-spin problems, and  $\ket{T_+}$ and $\ket{T_-}$ states become negligible in the dynamics.

Since conventional EPR experiments operate at fields above 10 \mT and over 1000 \mT, the effective tensor-network bond-dimension requirements in such regimes are substantially reduced. Scalability with respect to nuclear count is further discussed in Appendix \ref{subsec:scalability}.

\section{Anisotropic magnetosensitivity of ``two'' radical pairs}
\label{subsec:radicalhop}
\subsection{Background and physical setting}

The MPDO framework enables the study of electron hopping between radical pairs in cryptochrome, a candidate magnetosensor in the avian retina. In cryptochrome, cascaded chains of tryptophan (Trp) residues relay photo-generated charges \cite{giovaniLightinducedElectronTransfer2003, mullerDiscoveryFunctionalAnalysis2015, nohrDeterminationRadicalRadical2017, nohrExtendedElectronTransferAnimal2016,zoltowskiChemicalStructuralAnalysis2019,xuMagneticSensitivityCryptochrome2021}.

Upon photoexcitation of flavin adenine dinucleotide (FAD), an electron hole is transferred to \Trp{A} to form the initial radical pair \RP{A} and is then relayed to \Trp{B}, \Trp{C}, and \Trp{D} on an ultrafast timescale.
Fig.~\ref{fig:crypto}a schematically illustrates this process. Experiments indicate that \RP{C} and \RP{D} may coexist in equilibrium, while interconversion between \RP{A} $\rightarrow$ \RP{B} and \RP{B} $\rightarrow$ \RP{C} occurs ultrafast \cite{xuMagneticSensitivityCryptochrome2021}.

We simulate the simultaneous dynamics of the two radical pairs \RP{C} and \RP{D}, consisting of one flavin anion and two tryptophan cations. Migratory birds are hypothesised to detect direction using anisotropic magnetic modulation of spin-selective product yields in such transient radical pairs \cite{ritzMagneticCompassBirds2009, horeRadicalPairMechanismMagnetoreception2016}.
Since the geomagnetic field is extremely weak ($\sim 0.05\;\mT$), subtle anisotropic responses are difficult to detect experimentally. An exact quantum-mechanical simulation of such spin dynamics can, therefore, provide mechanistic insight into avian magnetoreception.
\subsection{Electron-hopping model and tensor-network construction}
We consider a system comprising two electronic spins and the nuclear spins from FAD, \Trp{C} and \Trp{D}, which we shall denote by $i=1,2,3$, respectively.
Nuclear spins of the flavin anion are shared between the two radical pairs. The electronic sector spans eight states $\ket{\uparrow\uparrow 0}$, $\ket{\uparrow\downarrow 0}$, $\ket{\downarrow\uparrow 0}$, $\ket{\downarrow\downarrow 0}$, $\ket{\uparrow 0 \uparrow}$, $\ket{\uparrow 0 \downarrow}$, $\ket{\downarrow 0 \uparrow}$, and $\ket{\downarrow 0 \downarrow}$.
which are rotated into the radical-pair basis $\ket{T_{+}^{\mathrm{C}}}$, $\ket{T_0^{\mathrm{C}}}$, $\ket{S^{\mathrm{C}}}$, $\ket{T_-^{\mathrm{C}}}$, $\ket{T_+^{\mathrm{D}}}$, $\ket{T_0^{\mathrm{D}}}$, $\ket{S^{\mathrm{D}}}$, and $\ket{T_-^{\mathrm{D}}}$ in the usual manner. In the vMPDO representation, physical sites are ordered as \Trp{C} nuclei $\to$ first half of FAD nuclei $\to$ electronic spins $\to$ the second half of FAD nuclei $\to$ \Trp{D} nuclei.
Within each molecule, nuclear sites are arranged such that strongly hyperfine-coupled spins are positioned closest to their electronic site in the tensor train. Hopping kinetic constants are taken from Ref.
\cite{xuMagneticSensitivityCryptochrome2021} ($k_{\mathrm{C}\to \mathrm{D}} = 15\;\nsinv$ and $k_{\mathrm{D}\to \mathrm{C}} = 13\;\nsinv$), which are three orders of magnitude faster than Haberkorn recombination rates $k_S^{W}=k_f^{W}+k_r^W, k_T^{W}=k_f^{W}$ for $W\in\{{\mathrm{C}}, \mathrm{D}\}$ where the singlet recombination rate $k_r^{\mathrm{C}} =17 \;\usinv$, $k_r^{\mathrm{D}} = 0.0 \;\usinv$ and proton transfer rate to a stabilised product $k_f^{\mathrm{C}} = 5.7 \;\usinv$, $k_f^{\mathrm{D}} = 10 \;\usinv$.
Assuming the electron spin does not flip during hopping, we describe hopping using Lindblad operators,
\begin{equation}
\begin{gathered}
  \mathcal{D}[\hat{\rho}] = \sum_{j\in\{\mathrm{C}\to\mathrm{D}, \mathrm{D}\to\mathrm{C}\}} \hat{L}_{j} \hat{\rho} \hat{L}_{j}^\dagger - \frac{1}{2} \hat{L}_{j}^\dagger \hat{L}_{j} \hat{\rho} - \frac{1}{2} \hat{\rho} \hat{L}_{j}^\dagger \hat{L}_{j}
  \\
  \hat{L}_{\mathrm{C}\to \mathrm{D}} = \sqrt{k_{\mathrm{C}\to \mathrm{D}}} \left( \ket{\mathrm{D}}\bra{\mathrm{C}}\right)
  \\
  \hat{L}_{\mathrm{D}\to \mathrm{C}} = \sqrt{k_{\mathrm{D}\to \mathrm{C}}} \left( \ket{\mathrm{C}}\bra{\mathrm{D}}\right),
\end{gathered}
\end{equation}
where $\ket{\mathrm{C}}\bra{\mathrm{D}} := \ket{S^\mathrm{C}}\bra{S^\mathrm{D}} + \ket{T_+^\mathrm{C}}\bra{T_+^\mathrm{D}} + \ket{T_-^\mathrm{C}}\bra{T_-^\mathrm{D}} + \ket{T_0^\mathrm{C}}\bra{T_0^\mathrm{D}}$, and $\ket{\mathrm{D}}\bra{\mathrm{C}}$ is the transpose of $\ket{\mathrm{C}}\bra{\mathrm{D}}$.
Alternative frameworks model hopping as (i) a reset of nuclear environments \cite{schultenSemiclassicalDescriptionElectron1978, gordonRotationalDiffusionMolecules1966}, or (ii) two density operators that exchange population \cite{kattnigElectronSpinRelaxation2016}.
In contrast, our approach retains nuclear memory and treats the coupled system within a single density operator with shared FAD nuclear sites.

In vMPDO form, the nominal electron-site dimension is $\left(d_{\mathrm{C}}+d_{\mathrm{D}}\right)^2=64$ where $d_{\mathrm{C}} =d_{\mathrm{D}} = 4$. Since \Trp{C} and \Trp{D} share FAD nuclei and there is no diabatic coupling between electronic sectors in the Liouvillian, all off-diagonal blocks $\ket{X^\mathrm{C}}\bra{Y^\mathrm{D}}$ and $\ket{X^\mathrm{D}}\bra{Y^\mathrm{C}}$ for $X,Y\in\{T_+, T_0, S, T_-\}$ may be projected out.
Consequently, the relevant electronic Liouville dimension reduces to $d_C^2+d_D^2=32$.

A further reduction applies to the three equivalent methyl protons in FAD. The tensor product $(\tfrac{1}{2})^{\otimes 3}$ decomposes into $\left(I_{\mathrm{tot}}=\tfrac{3}{2}\right) \oplus 2\times \left(I_{\mathrm{tot}}=\tfrac{1}{2}\right)$, with sector dimensions $4$ and $2$ (twice), respectively.
The details of the symmetry reduction technique are described in Appendix \ref{subsec:symmetry-reduction}.
Off-diagonal blocks between different nuclear-multiplicity sectors may be projected out, reducing the physical dimension of methyl-hydrogen sites from $(2^3)^2=64$ to $4^2+2^2=20$.
Equivalent reductions can also be implemented via explicit quantum-number conservation in tensor networks.

Exchange couplings $J$ are taken from out-of-phase ESEEM measurements \cite{gravellSpectroscopicCharacterizationRadical2025}; dipolar couplings $\mathbf{D}$ are computed from PDB 6PU0\cite{zoltowskiChemicalStructuralAnalysis2019} using a point-dipole model; hyperfine tensors $\mathbf{A}$ are obtained from density functional theory \cite{neeseORCAProgramSystem2012}.
The resulting anisotropic coefficients $\mathbf{D}_{1i}-2J_{1i}\mathbb{1}_3\;(i\in2,3)$ and $\mathbf{A}_{ij}\;(i\in\{1, 2, 3\})$ are listed in Appendix \ref{subsec:aniso-params}.

\subsection{Anisotropic spin dynamics}
\begin{figure*}
    \centering
  \begin{overpic}[width=.50\linewidth]{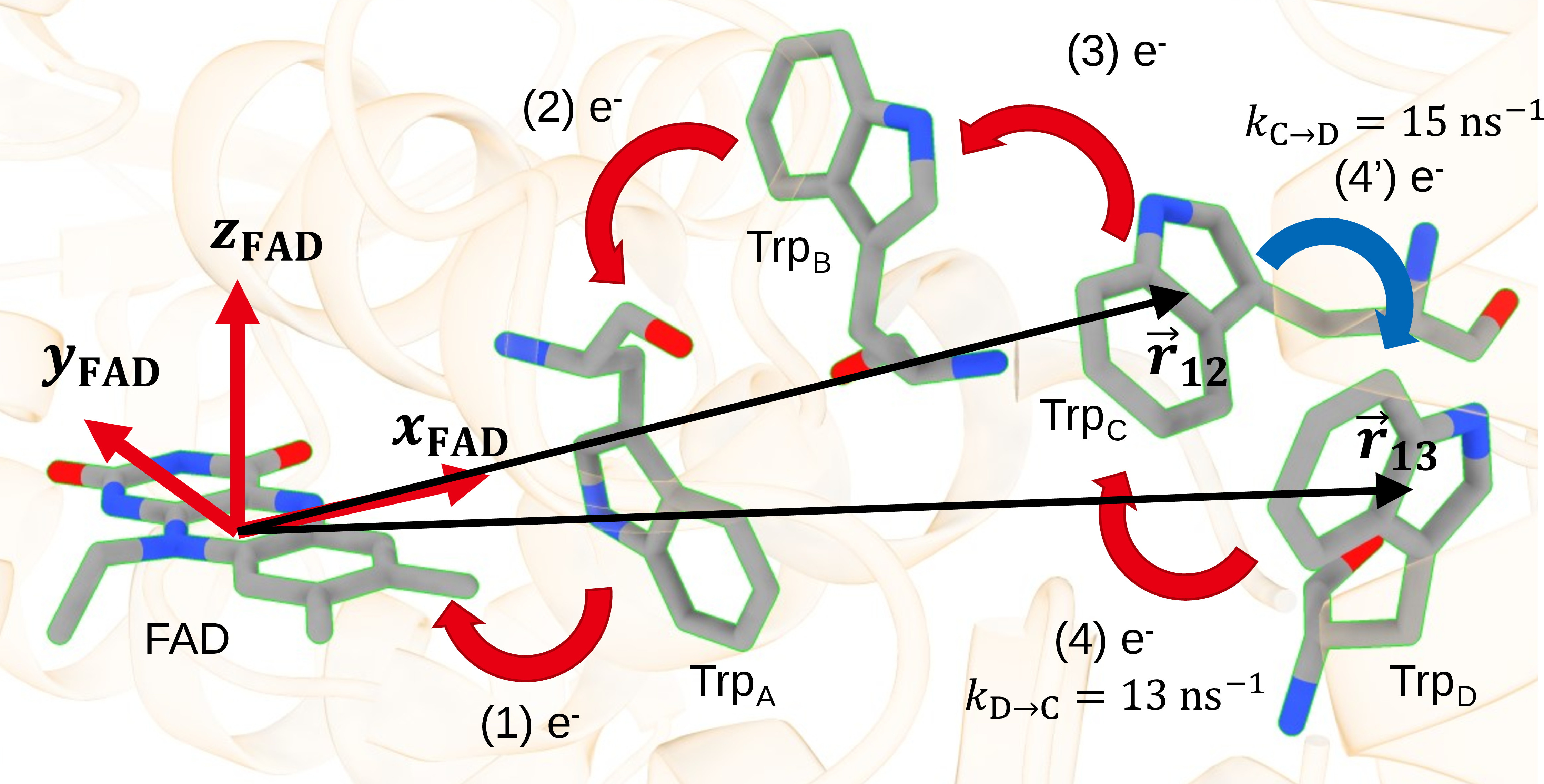}
    \put(1,50){\textbf{a}} % x=5 %, y=4 % from bottom-left
  \end{overpic}
  \begin{overpic}[width=.44\linewidth]{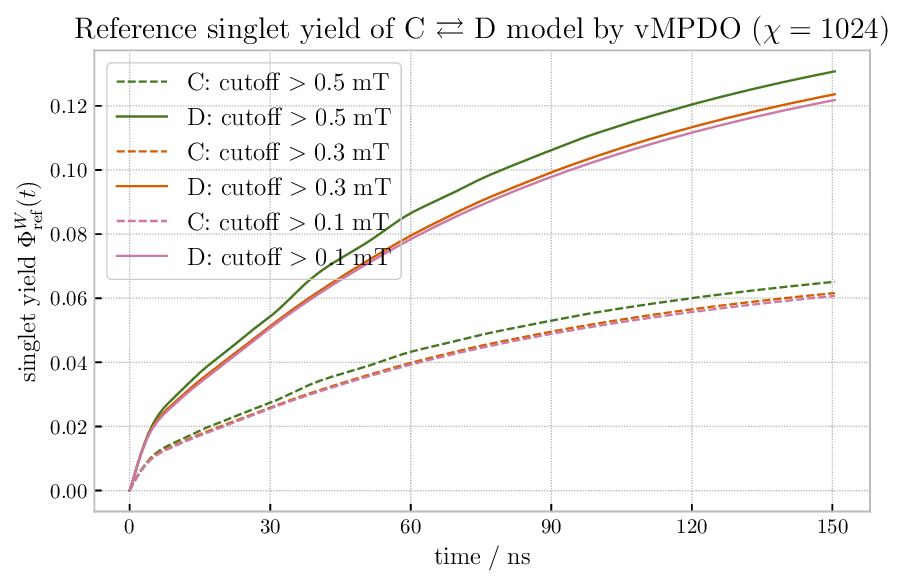}
    \put(1,60){\textbf{b}} % x=5 %, y=4 % from bottom-left
  \end{overpic}
  \newline
  \newline
  \begin{overpic}[width=0.49\linewidth]{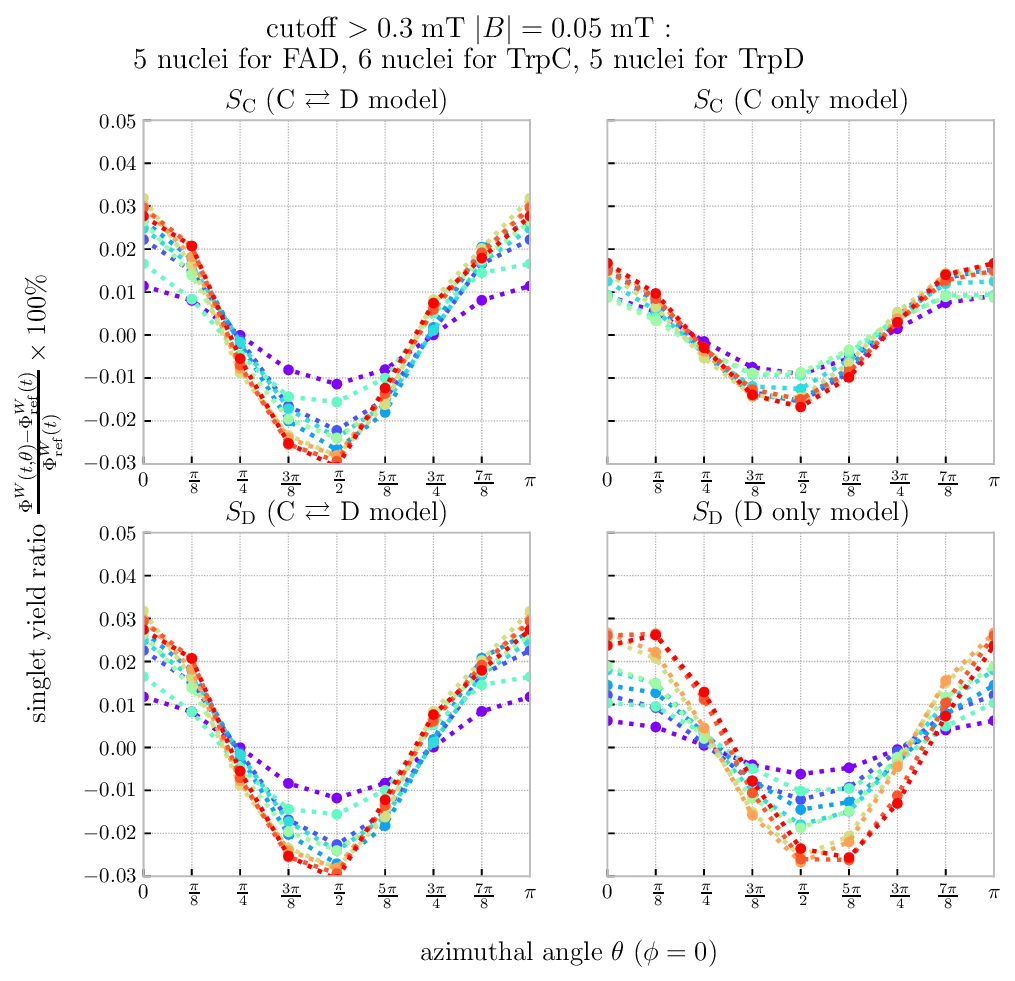}
    \put(1,90){\textbf{c}} % x=5 %, y=4 % from bottom-left
  \end{overpic}
  \begin{overpic}[width=0.49\linewidth]{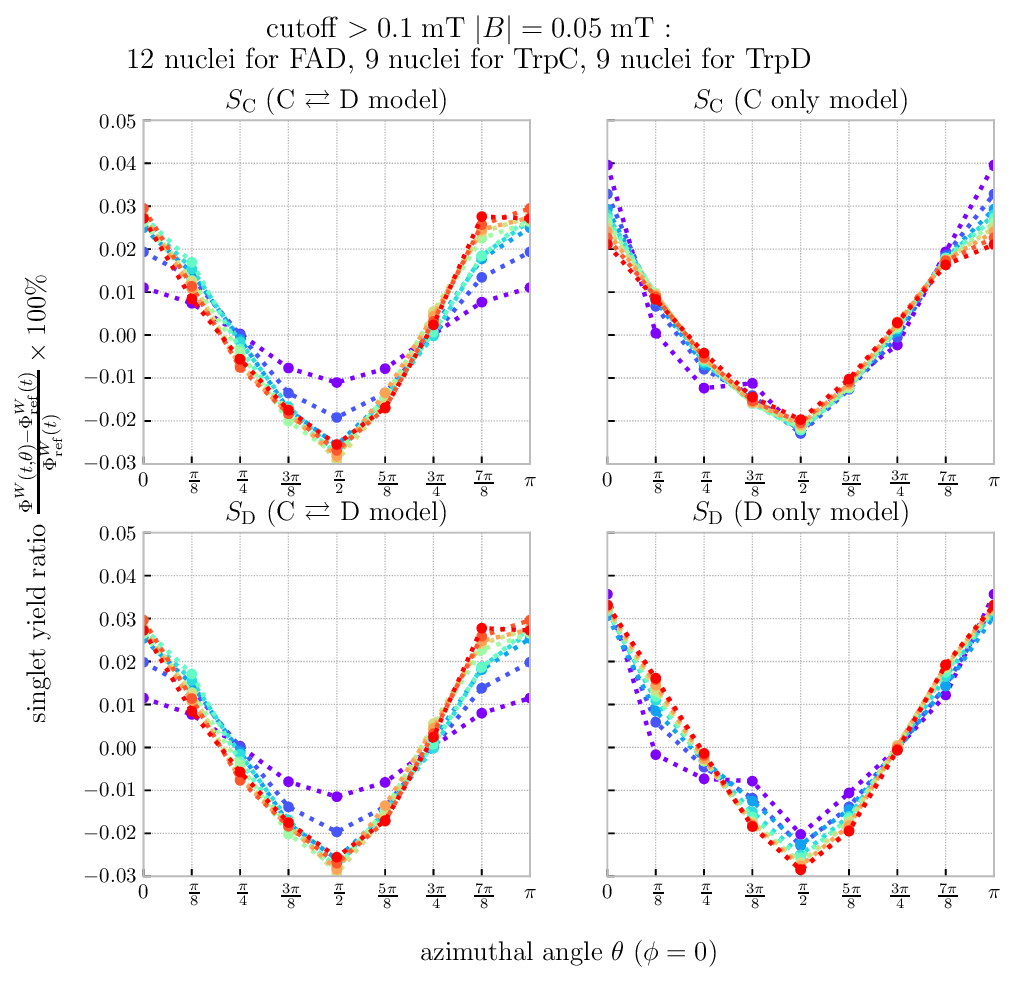}
    \put(1,90){\textbf{d}} % x=5 %, y=4 % from bottom-left
  \end{overpic}
  \begin{overpic}[width=0.49\linewidth]{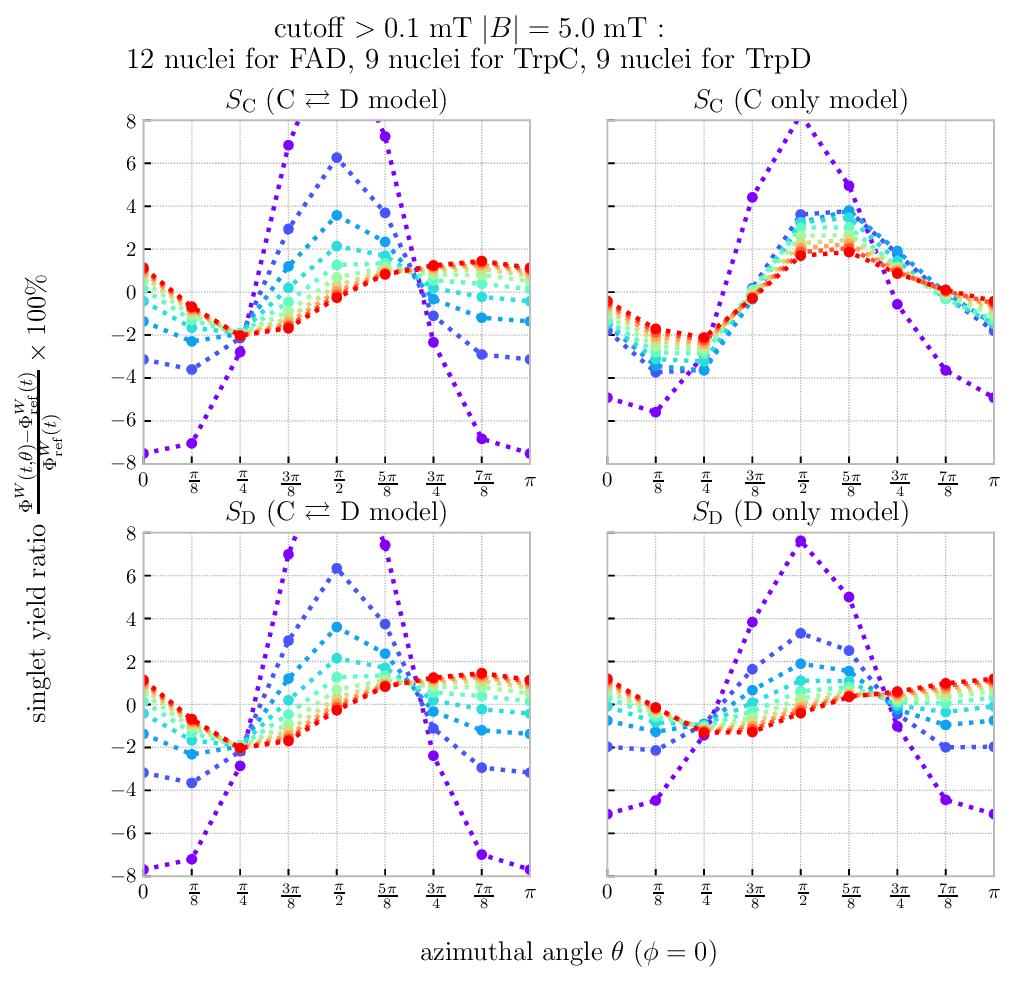}
    \put(1,90){\textbf{e}} % x=5 %, y=4 % from bottom-left
  \end{overpic}
  \begin{overpic}[width=0.19\linewidth]{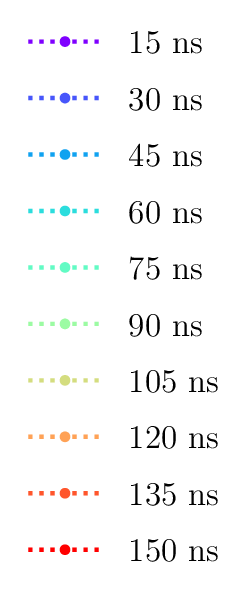}
    %\put(1,90){\textbf{e}} % x=5 %, y=4 % from bottom-left
  \end{overpic}
    \caption{Electron hopping geometry and anisotropic singlet yields.
    \textbf{(a)} Sequential electron transfer from \Trp{A} to \Trp{D} and relative orientations of aromatic backbones in the crystal. Spin bases are defined in the FAD frame, with the $z$-axis normal to the isoalloxazine plane and right-handed hyperfine tensors.
    %Vectors $\vec{r}_{12}$ and $\vec{r}_{13}$ connect the FAD isoalloxazine centre‐of‐mass to those of \Trp{C} and \Trp{D}, respectively.
    %
    \textbf{(b)} Cumulative singlet yield $\Phi^W_{\mathrm{ref}}(t)$ for different hyperfine-cutoff thresholds. Included nuclear counts are
    $\left(N_{\mathrm{FAD}}, N_{\mathrm{C}}, N_\mathrm{D}\right) = (1, 3, 3)$ for $\mathrm{cutoff} > 0.5 \;\mT$,
    $\left(N_{\mathrm{FAD}}, N_{\mathrm{C}}, N_\mathrm{D}\right) = (5, 6, 5)$ for $\mathrm{cutoff} > 0.3 \;\mT$,
    and
    $\left(N_{\mathrm{FAD}}, N_{\mathrm{C}}, N_\mathrm{D}\right) = (12, 9, 9)$ for $\mathrm{cutoff} > 0.1 \;\mT$.
    \textbf{(c-e)}
    Transient singlet-yield ratios vs. magnetic-field azimuth $\theta$, referenced to the azimuthal average. Panel (c) uses cutoff > $0.3\;\mT$.
    Panel (d) uses cutoff > $0.1\;\mT$.
    Panel (e) uses cutoff > $0.1\;\mT$ and strong field $|B| = 5.0 \;\mT$.
    }
    \label{fig:crypto}
\end{figure*}

To quantify the dependence of spin dynamics on magnetic-field orientation, we compute the relative singlet-yield ratio
\begin{equation}
\begin{gathered}
    M^W(t, \theta_k) = \frac{\Phi^W(t, \theta_k) - \Phi^W_{\mathrm{ref}}(t)}{\Phi^W_{\mathrm{ref}}(t)},
\\
\Phi^W(t, \theta)=\int_0^t \mathrm{d}\tau \; k_f^W\mathrm{Tr}\left[\hat{P}_S^W\hat{\rho}(\tau, \theta)\right],
    \\\quad \theta_k\in\left\{\frac{\pi k}{8} \middle| k=0, 1, 2,\ldots,7\right\},
    \\\quad W\in\{\mathrm{C}, \mathrm{D}\},
\end{gathered}
\end{equation}
where the density operator $\hat{\rho}(t, \theta)$ is propagated under a magnetic field $\mathbf{B}=\left[B_0\sin\theta, 0, B_0\cos\theta\right]^\top$ with $B_0=0.05\;\mT$ (Fig.~\ref{fig:crypto}c, d) and $B_0=5.0\;\mT$ (Fig.~\ref{fig:crypto}e).
The reference yield is the azimuthal average $\Phi^W_{\mathrm{ref}}(t) = \frac{1}{8}\sum_{k=0}^7 \Phi^W(t, \theta_k)$ with the polar angle fixed at $\phi=0$.
The initial state is the singlet of \RP{C} and $\hat{\rho}=\hat{P}_S^{\mathrm{C}}$; due to hopping, equilibrium with \RP{D} is reached at a population ratio of $k_\mathrm{D\to C}:k_\mathrm{C \to D}=1.3:1.5$.
All simulations use bond dimensions $\chi=r=1024$ and an Arnoldi integrator.
%
% Although it is not employed in this work, we considered a locally purified approach for simulating the Lindblad master equation  \cite{wernerPositiveTensorNetwork2016}, which is written in \ref{subsec:lindblad-tn}.

To examine the role of anisotropic hyperfine couplings, we vary a threshold (``cutoff'') defined as the mean absolute eigenvalue of the hyperfine tensor. Reference singlet yields are shown in Fig~\ref{fig:crypto}b.
A $0.3\;\mT$ cutoff (16 nuclei) suffices for approximate convergence. At geomagnetic field strength, Fig.~\ref{fig:crypto}c, d shows that a $z$-aligned field enhances the singlet-yield ratio by 0.10\%.

We further compare simulations of (i) independent \RP{C} and \RP{D}, and (ii) the coupled two-pair system. The anisotropic response of \RP{D} sharpens over time, whereas \RP{C} does not; the two-pair model reproduces the \RP{D}-like behaviour.
When too few hyperfine nuclei are included, \RP{D} exhibits a distinct time-dependence in its anisotropy, indicating that nuclei excluded at the 0.3 \mT threshold (e.g., aromatic nitrogens in \Trp{D}) materially influence the anisotropic response. Inclusion of the composite radical-pair model changes the time-dependence slightly, but not the gross orientation behaviour.

At $5.0\;\mT$, Fig.~\ref{fig:crypto}e shows that the anisotropic amplitude increases by about two orders of magnitude; however, both the angular dependence and time-dependence differ qualitatively from the geomagnetic regime. This follows from the distinct mechanisms governing spin sensitivity: hyperfine/low-field effects at geomagnetic scales versus Zeeman-dominated responses at high fields.

\section*{Practical considerations and future improvements}
Stochastic MPS, while trajectory-intensive, is most economical in bond dimension and is well-suited for simulations with stochastic time-dependent Hamiltonians (e.g., coupled to molecular dynamics \cite{pazeraSpinDynamicsRadical2024, benjaminMagnetosensitivityModelFlavin2025}). When GPU acceleration is available, deterministic LPMPS and vMPDO are advantageous for parameter scans over magnetic fields, providing short wall-clock times. For incorporating additional relaxation channels \textemdash \ e.g., Bloch-Redfield relaxation or Lindblad jumps \textemdash \ vMPDO provides the most practical framework.

Although tensor networks enable efficient sub-microsecond simulations, extending beyond the Larmor period in geomagnetic fields ($\sim$700 ns) remains challenging because bond dimensions grow rapidly. Future progress may come from Clifford disentanglers\cite{melloCliffordDressedTimeDependent2025, qianCliffordCircuitsAugmented2025}, which suppress entanglement growth and thereby enable long-time propagation at modest bond dimensions.
Incorporating more accurate ab initio evaluations of exchange and dipolar couplings, dynamic structural input from molecular-dynamics trajectories, and additional relaxation channels (e.g., singlet-triplet dephasing \cite{shushinEffectSpinExchange1991, golesworthySingletTripletDephasing2023} and random field relaxation \cite{kattnigElectronSpinRelaxation2016}) will further enhance predictive capability.

\begin{acknowledgments}
We thank Peter J. Hore (University of Oxford), Christiane R. Timmel (University of Oxford) and Claudia E. Tait (University of Oxford) for valuable discussions.
KH was supported by JSPS KAKENHI (Grant No. JP23KJ1334).
YK was supported by JSPS KAKENHI (Grant No. JP23H01921), JST FOREST Program (Grant No. JPMJFR221R), JST CREST Program (Grant No. JPMJCR23I6), JST MEXT Q-LEAP Program (Grant No. JPMXS0120319794) and JST ASPIRE Program (Grant No. JPMJAP2423).
LMA was supported by the JST ASPIRE Program (Grant No. JPMJAP2423).
DF was supported by a Clarendon Scholarship from the University of Oxford.
Part of computation was performed using the Research Center for Computational Science, Okazaki, Japan (Project: 25-IMS-C029).
\end{acknowledgments}

\section*{Conflict of Interest Statement}
The authors have no conflicts to disclose.

\section*{Data Availability Statement}
The stochastic MPS, vMPDO, LPMPS, MPO construction, and symmetry reduction are implemented in Python; SW and SC are implemented in Julia, and the stochastic full wavefunction is implemented in Fortran.
The guide and scripts are all available at \url{https://github.com/KenHino/radicalpair-tensornetwork}.
Tensor network methods are also available in the \texttt{RadicalPy} library~\cite{antillRadicalPyToolSpin2024}.

\appendix
\section{Tensor network methods}
\label{sec:methods}
\subsection*{Stochastic matrix product states}
We represent the radical-pair wavefunction in a matrix-product-state (MPS) form and site basis ordering
\begin{widetext}
\begin{equation}
    \begin{aligned}
  \Ket{\Psi} &= \sum_{\sigma_1, \sigma_2, \cdots, \sigma_N} \sum_{\alpha_1, \cdots, \alpha_{N-1}} A\substack{\sigma_1\\\alpha_1} A\substack{\sigma_2\\\alpha_1\alpha_2} \cdots A\substack{\sigma_N\\\alpha_{N-1}}
  \ket{\sigma_1, \sigma_2, \cdots, \sigma_N}
  \\
  \ket{\sigma_k} &=
  \begin{cases}
    \ket{\sigma_{\mathrm{nuc}}^{(1, k)}} & \text{for } k=1,2,\cdots,N_1 \\
    \ket{\sigma_{\mathrm{el}}} & \text{for } k=N_1+1 \\
    \ket{\sigma_{\mathrm{nuc}}^{(2, k-N_1-1)}} & \text{for } k=N_1+2,N_1+3,\cdots,N_1+N_2+1,
  \end{cases}
  \end{aligned}
\end{equation}
\end{widetext}
and each core tensor $A\substack{\sigma_k \\ \alpha_{k-1}\alpha_k} \in \mathbb{C}^{m_{k-1} \times d_k \times m_k}$ carries a bond dimension $m_k$ that controls accuracy via the virtual indices $\alpha_k=1,2,\cdots,m_k$.
Physical dimensions are $d_k=4$ for the two-electron site and $d_k=2I^{(i,j)}+1$ for a nuclear site. This reduces storage from $\mathcal{O}(d^N)$ to $\mathcal{O}(Ndm^2)$.
A schematic diagram of the MPS approach is depicted in Fig.~\ref{fig:mps-mpo}a.
For the Monte-Carlo evaluation of Eq.~(\ref{eq:pop-observable}), we employ spin-coherent state sampling \cite{faySpinRelaxationRadical2021}(explicit formula in Appendix \ref{subsec:Kdeps}). To evolve with an MPS-compatible operator, we construct a matrix-product operator (MPO) automatically using a symbolic scheme based on bipartite-graph matching \cite{renGeneralAutomaticMethod2020a}; the MPO diagram is in Fig.~\ref{fig:mps-mpo}a, and the explicit radical-pair MPO form is given in the MPO subsection below.

\subsection*{Vectorised matrix product density operator}
To deterministically evaluate Eq.~(\ref{eq:pop-observable}), we use the matrix-product density-operator (MPDO) formalism in two flavours: (i) direct vectorisation (vMPDO) and (ii) a locally purified representation (LPMPS)\cite{verstraeteMatrixProductDensity2004a,cuevasPurificationsMultipartiteStates2013,wernerPositiveTensorNetwork2016}.
Because Liouville space scales as the square of the Hilbert dimension, MPDOs are more demanding and require superoperator propagation techniques.
An MPDO with an MPO-like structure is
\begin{widetext}
\begin{equation}
  \hat{\rho} =
  \sum_{\substack{\sigma_1, \sigma_2, \cdots, \sigma_N \\ \sigma_1^\prime, \sigma_2^\prime, \cdots, \sigma_N^\prime}}
  \sum_{\gamma_1, \cdots, \gamma_{N-1}}
  C\substack{\sigma_1^\prime\\\gamma_1\\\sigma_1}
  C\substack{\sigma_2^\prime\\\gamma_1\gamma_2\\\sigma_2}
  \cdots
  C\substack{\sigma_N^\prime\\\gamma_{N-1}\\\sigma_N}
  \ket{\sigma_1^\prime, \sigma_2^\prime, \cdots, \sigma_N^\prime}
  \bra{\sigma_1, \sigma_2, \cdots, \sigma_N},
\end{equation}
\end{widetext}
where $\gamma_k = 1, 2, \cdots, \chi_k$ and  $\chi_k$ are the MPDO bond dimensions. Storage reduces from $\mathcal{O}(d^{2N})$ to $\mathcal{O}(N d^2 \chi^2)$.
Vectorising via $\mathbb{C}^{4Z\times 4Z}$ to $\mathbb{C}^{(4Z)^2}$ yields
$
  \mathrm{vec}(\hat{\rho}) =: \dket{\rho}
$, the ``twin-space''\cite{borrelliDensityMatrixDynamics2019a}.
\begin{equation}
\begin{aligned}
  &\dket{\rho}= \\
  &
  \sum_{\rho_1, \rho_2, \cdots, \rho_N}
  \sum_{\gamma_1, \cdots, \gamma_{N-1}}
  C\substack{\rho_1\\\gamma_1}
  C\substack{\rho_2\\\gamma_1\gamma_2}
  \cdots
  C\substack{\rho_N\\\gamma_{N-1}}
  \dket{\rho_1, \rho_2, \cdots, \rho_N},
\end{aligned}
\end{equation}
where
$\dket{\rho_k} := \mathrm{vec}(\ket{\sigma_k^\prime}\bra{\sigma_k}) \in \mathbb{C}^{d_k^2}$
is the vectorised single-body physical site.
We shall call this formulation vectorised MPDO (vMPDO).
Since the superoperators in the Liouville-von Neumann equation are encoded as linear operators using the identity,
$
\mathrm{vec}\left(A\rho B\right) = \left(B^\top \otimes A\right) \mathrm{vec}(\rho),
$
The Liouvillian inherits a sum-of-products structure, enabling the symbolic MPO construction of the superoperator. The vMPDO diagram is shown in Fig~\ref{fig:mps-mpo}a.
Physical consistency requires Hermiticity, positivity, and trace preservation of $\hat{\rho}$.
However, a finite-$\chi$ vMPDO approximates $\hat{\rho}$ within a tensor-train manifold that does not enforce these constraints and retains redundant components (e.g., imaginary parts of diagonals and lower-triangular entries). Insufficient $\chi$ can therefore violate complete positivity or trace conservation.

\subsection*{Locally purified matrix product states}
To enforce physicality, we employ locally purified MPS (LPMPS), where the physical density operator is the partial trace over ancilla sites interleaved with each physical site:
\begin{widetext}
\begin{equation}
\begin{gathered}
   \hat{\rho}_{\mathrm{phys}}(t) = \mathrm{Tr}_{\{s_1, s_2, \cdots, s_N\}} \left\{\ket{\Psi(t)}\bra{\Psi(t)}\right\}
   \\
  \Ket{\Psi} = \sum_{
    \substack{\sigma_1, \sigma_2, \cdots, \sigma_N \\ s_1, s_2, \cdots, s_N}}
    \sum_{\alpha_1, \alpha_2, \cdots, \alpha_{2N-1}}
    A\substack{\sigma_1\\\alpha_1}
    A\substack{s_1\\\alpha_1\alpha_2}
    A\substack{\sigma_2\\\alpha_2\alpha_3}
    \cdots
    A\substack{s_N\\\alpha_{2N-1}}
  \ket{\sigma_1, s_1, \sigma_2, \cdots, s_N}.
\end{gathered}
\end{equation}
\end{widetext}
We denote the LPMPS bond dimension by $r\;(\alpha_k=1,2,\cdots,r_k)$. The unitary part evolves under $\hat{H} \otimes \mathbb{1}_{\mathrm{anc}}$.
For the standard mixed initial state
\begin{equation}
  %\hat{\rho}_{\mathrm{phys}}(0)
  %$ =
  \ket{S}\bra{S} \otimes \bigotimes_{i=1}^{2}\bigotimes_{j=1}^{N_i}
  \left\{\frac{1}{2I_{i,j}+1}\sum_{m=-I_{i,j}}^{I_{i,j}}\ket{I_{i,j},m}\bra{I_{i,j},m}\right\},
\end{equation}
the purified initial state can be taken as
\begin{equation}
  %\ket{\Psi(0)} =
  \ket{S}_{\mathrm{phys}}
  \otimes
  \bigotimes_{i=1}^{2}
  \bigotimes_{j=1}^{N_i}
  \left\{
  \frac{
  \sum_{m=-I_{i,j}}^{I_{i,j}} \ket{I_{i,j}, m}_{\mathrm{phys}} \ket{I_{i,j}, m}_{\mathrm{anc}}
  }
  {\sqrt{2I_{i,j}+1}}
  \right\}.
\end{equation}
Thus, the initial physical-ancilla bond on a nuclear site is $2I_{i,j}+1$, while bonds between different physical–ancilla sites start at 1.
Since the electronic state is pure (singlet), we omit electronic ancillas; the total site count is $1+2N_1+2N_2$.
A diagram is given in Fig~\ref{fig:mps-mpo}a.

The time evolution of MPS, vMPDO, and LPMPS is performed using the time-dependent variational principle (TDVP).
Fig.~\ref{fig:mps-mpo}b illustrates propagation on the low-rank manifold via tangent-space projection.

\subsection*{Time-dependent variational principle (TDVP) for tensor-network states}
\label{subsec:tdvp}
With vectorised MPDOs and an MPO Liouvillian, propagation follows the same TDVP machinery as for MPS/LPMPS.
The detailed routine of the time evolution of the tensor train is described in references \cite{lubichTimeIntegrationTensor2015a,haegemanUnifyingTimeEvolution2016b}.
Here we briefly explain the idea.
Let
\begin{equation}
    \frac{\mathrm{d}}{\mathrm{d}t}\ket{X} = \hat{O}\ket{X}
\end{equation}
where $\hat{O}$ is the MPO of $-i\hat{H}$ or the Liouvillian $\dhat{\mathcal{L}}$, and $\ket{X}$ (MPS or MPDO) lies on a low-rank manifold $\mathcal{M}$.
The Dirac-Frenkel principle gives
\begin{equation}
    \Braket{\delta X|\frac{\mathrm{d}}{\mathrm{d}t} - \hat{O}|X} = 0.
\end{equation}
Using the projection operator $\hat{\mathcal{P}}_{T_{X(t)}\mathcal{M}}$ onto the tangent space at $X(t)$, we can rewrite the time-dependent equation as
\begin{equation}
    \ket{X(t+\Delta t)} = \exp\left(\hat{\mathcal{P}}_{T_{X(t)}\mathcal{M}}\hat{O}\Delta t\right)\ket{X(t)}.
\end{equation}
Assuming the tensor train $\ket{X}$ is written as
\begin{equation}
    \sum_{\sigma_1,\cdots,\sigma_N}
    U^{\sigma_1}_{1}\cdots U^{\sigma_{j-1}}_{j-1}\Psi_{j}^{\sigma_j}V^{\sigma_j}_{j+1}\cdots V^{\sigma_N}_{N}
    \ket{\sigma_1,\cdots,\sigma_N}
\end{equation}
where the orthogonality conditions $\sum_{\sigma_j} (U_j^{\sigma_j})^{\dagger}U_j^{\sigma_j} = \mathbb{1}$ and $\sum_{\sigma_j} V_j^{\sigma_j}\left(V_j^{\sigma_j}\right)^\dagger = \mathbb{1}$ hold, the projection operator $\hat{\mathcal{P}}_{T_{X(t)}\mathcal{M}}$ is given by
\begin{equation}
    \hat{\mathcal{P}}_{T_{X(t)}\mathcal{M}} =
    \sum_{j=1}^{N}
    \hat{\mathcal{P}}_{L}^{[1:j-1]} \otimes \hat{\mathbb{1}}_j \otimes \hat{\mathcal{P}}_{R}^{[j+1:N]}
    -
    \sum_{j=1}^{N-1}
    \hat{\mathcal{P}}_{L}^{[1:j]} \otimes \hat{\mathcal{P}}_{R}^{[j+1:N]}
\end{equation}
where
\begin{widetext}
\begin{equation}
    \begin{gathered}
        \hat{\mathcal{P}}_{L}^{[1:j]} =
        \sum_{\substack{\sigma_1^\prime,\cdots,\sigma_{j}^\prime\\
        \sigma_1,\cdots,\sigma_{j}}} U^{\sigma_1^\prime}_{1}\cdots U^{\sigma_{j}^\prime}_{j} \ket{\sigma_1^\prime,\cdots,\sigma_{j}^\prime} \bra{\sigma_1,\cdots,\sigma_{j}} \left(U^{\sigma_1}_{1}\cdots U^{\sigma_{j}}_{j}\right)^\ast,
        \\
        \hat{\mathcal{P}}_{R}^{[j:N]} = \sum_{\substack{\sigma_{j}^\prime,\cdots,\sigma_N^\prime\\
        \sigma_{j},\cdots,\sigma_N}} V^{\sigma_{j}^\prime}_{j}\cdots V^{\sigma_N^\prime}_{N} \ket{\sigma_{j}^\prime,\cdots,\sigma_N^\prime} \bra{\sigma_{j},\cdots,\sigma_N} \left(V^{\sigma_{j}}_{j}\cdots V^{\sigma_N}_{N}\right)^\ast,
        \\
        \hat{\mathbb{1}}_j = \sum_{\sigma_j} \ket{\sigma_j} \bra{\sigma_j}.
    \end{gathered}
\end{equation}
\end{widetext}
The time evolution of the tensor train $\ket{X}$ is achieved through the Trotter decomposition of the propagator
\begin{equation}
    \begin{aligned}
    &\exp\left(\hat{\mathcal{P}}_{T_{X(t)}\mathcal{M}}\hat{O}\Delta t\right)
    \\
    &\simeq
    e^{\hat{\mathcal{P}}_{R}^{[2:N]}\hat{O}\frac{\Delta t}{2}}
    e^{-\hat{\mathcal{P}}_{L}^{[1:1]}\hat{\mathcal{P}}_{R}^{[2:N]}\hat{O}\frac{\Delta t}{2}}
    %e^{\hat{\mathcal{P}}_{L}^{[1:1]}\hat{\mathcal{P}}_{R}^{[3:N]}\hat{O}\frac{\Delta t}{2}}
    %e^{-\hat{\mathcal{P}}_{L}^{[1:2]}\hat{\mathcal{P}}_{R}^{[3:N]}\hat{O}\frac{\Delta t}{2}}
    \cdots
    e^{\hat{\mathcal{P}}_{L}^{[1:N-1]}\hat{O}\frac{\Delta t}{2}}
    \\
    &\hspace{1em}
    e^{\hat{\mathcal{P}}_{L}^{[1:N-1]}\hat{O}\frac{\Delta t}{2}}
    e^{-\hat{\mathcal{P}}_{L}^{[1:N-1]}\hat{\mathcal{P}}_{R}^{[N:N]}\hat{O}\frac{\Delta t}{2}}
    \cdots
    %e^{-\hat{\mathcal{P}}_{L}^{[1:1]}\hat{\mathcal{P}}_{R}^{[2:N]}\hat{O}\frac{\Delta t}{2}}
    e^{\hat{\mathcal{P}}_{R}^{[2:N]}\hat{O}\frac{\Delta t}{2}}
    + \mathcal{O}(\Delta t^3).
    \end{aligned}
\end{equation}
The core tensor of the tensor train is updated through a sweeping routine that proceeds from left to right and then from right to left.
\begin{figure}
  \centering
  \includegraphics[width=0.98\linewidth]{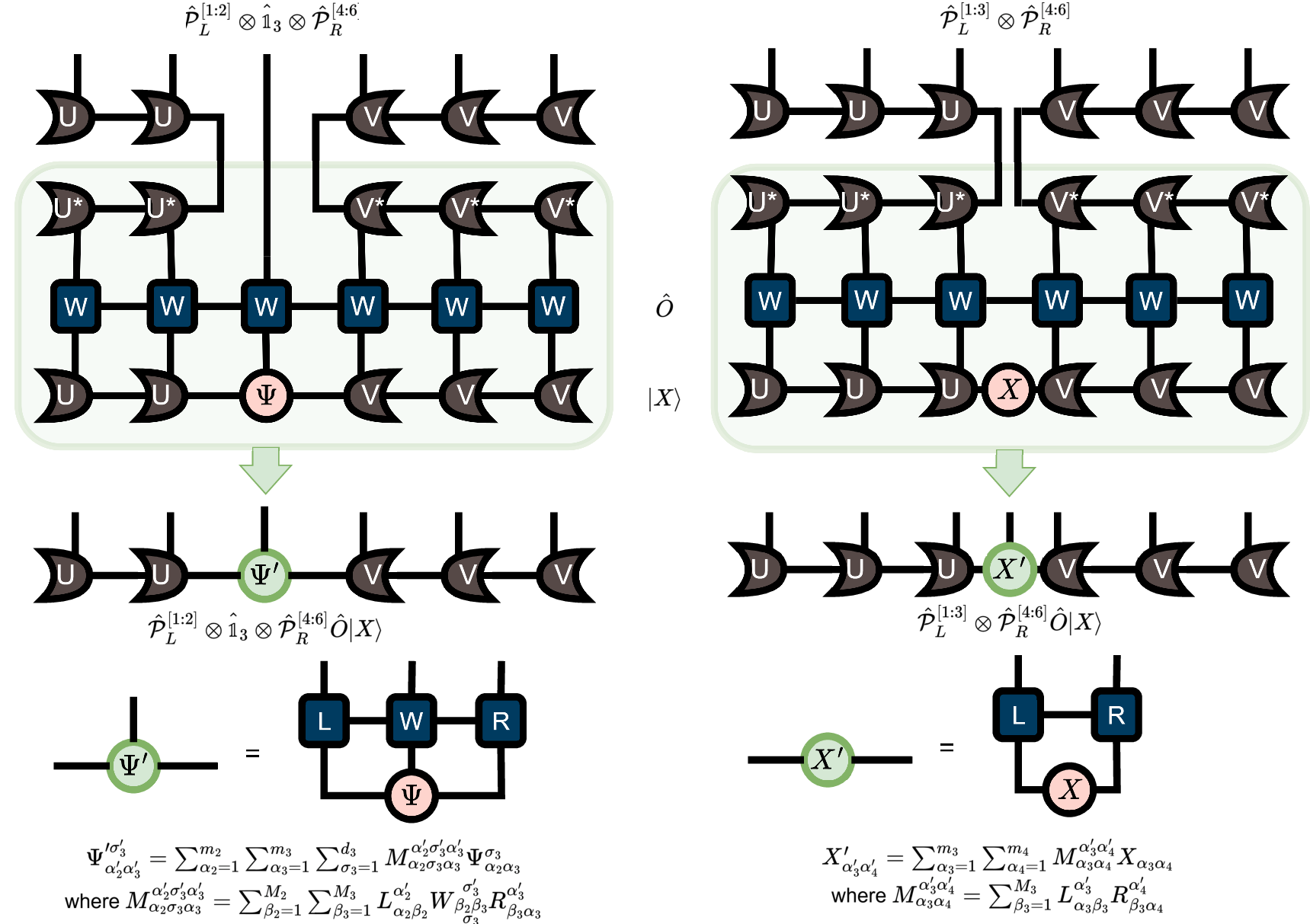}
  \caption{Tensor network diagram of the operation $\hat{\mathcal{P}}_L^{[1:2]}\otimes\hat{\mathbb{1}}_3\otimes\hat{\mathcal{P}}_R^{[4:6]}\hat{O}$ and
  $\hat{\mathcal{P}}_L^{[1:3]}\otimes\hat{\mathcal{P}}_R^{[4:6]}\hat{O}$
  on MPS $\ket{X}$ to obtain $\ket{X^\prime}$.}
  \label{fig:tdvp}
\end{figure}
As illustrated in Fig.~\ref{fig:tdvp}, the projection operator enables the time evolution of orthogonal centre tensors $\Psi^{\sigma_j}_{\alpha_{j-1},\alpha_j} \in \mathbb{C}^{m_{j-1} \times d_j \times m_j}$ and $X_j\in\mathbb{C}^{m_{j-1}\times m_j}$.
The iterative Krylov subspace method can be employed \cite{parkUnitaryQuantumTime1986a}, which is the most computationally expensive part of the algorithm and requires $\mathcal{O}(m^3d^2)$ operations for each effective matrix-vector multiplication in the Krylov iteration.
Once the orthogonal centre tensors are updated, the tensor is decomposed using QR (LQ) decomposition to obtain the next orthogonal centre tensors
\begin{equation}
  \Psi^{\sigma_j}_j = U_j^{\sigma_j} X_{j} = X_{j-1}V_j^{\sigma_j}.
\end{equation}
GPU acceleration substantially speeds both Krylov matrix-vector multiplication and local factorizations. For non-Hermitian evolution, we use an Arnoldi integrator.
While the method conserves $\braket{X|X}$ and the expectation value $\braket{X|\hat{O}|X}$ for Hermitian $\hat{O}$, the vMPDO trace $\mathrm{Tr}(\rho)=\langle\langle1|\rho\rangle\rangle$ is not guaranteed to be preserved.
%
%We have discussed other approaches to incorporate Lindblad master equation with tensor network method in \ref{subsec:lindblad-tn}.
%
\subsection*{MPO representation for radical pair Hamiltonian}\label{subsec:concrete-mpo}
An MPO enables polynomial-time application of $\hat{H}$ to an MPS:
\begin{equation}
  \hat{H} = \sum_{\beta_1, \cdots, \beta_{N-1}}
  \hat{W}\substack{\sigma_1^\prime\\\beta_1\\\sigma_1}
  \hat{W}\substack{\sigma_2^\prime\\\beta_1\beta_2\\\sigma_2}
  \cdots
  \hat{W}\substack{\sigma_N^\prime\\\beta_{N-1}\\\sigma_N}
\end{equation}
with core operators $\hat{W}\substack{\sigma_i^\prime\\\beta_{i-1}\beta_i\\\sigma_i} \in \mathbb{C}^{M_{i-1} \times d_i \times d_i \times M_i}$ and MPO bond $M_i$. We construct $\hat{H}$ using a symbolic bipartite-graph-matching method \cite{renGeneralAutomaticMethod2020a}, which avoids numerical SVD errors and is convenient for time-dependent Hamiltonians (pulsed EPR fields; MD-driven dipolar/exchange/hyperfine variations). Time-dependent MPOs are updated by modifying the relevant entries, whereas SVD-based constructions
\cite{hubigGenericConstructionEfficient2017a} would require recomputing the SVD at each step. For our Hamiltonian, an analytical MPO with a maximum bond dimension $M=5$ is obtained.
For illustration (electrons $i=1,2$; nuclei $j=1,2,3$), omitting $\gamma^{(\mathrm{e})}$ and operator subscripts, the first three nuclear cores are
\begin{equation}
  \hat{W}_1 =
  \begin{bmatrix}
    \hat{I}_x & \hat{I}_y & \hat{I}_z & 1
  \end{bmatrix}.
\end{equation}
The second core tensor, corresponding to the $(i,j)=(1,2)$ nucleus, is given by
\begin{equation}
  \hat{W}_2 =
  \begin{bmatrix}
    A_{xx}^{(1, 1)} & A_{yx}^{(1, 1)} & A_{zx}^{(1, 1)} & -B_x\gamma^{(\mathrm{n})}_{1,1} & 0 \\
    A_{xy}^{(1, 1)} & A_{yy}^{(1, 1)} & A_{zy}^{(1, 1)} & -B_y\gamma^{(\mathrm{n})}_{1,1} & 0 \\
    A_{xz}^{(1, 1)} & A_{yz}^{(1, 1)} & A_{zz}^{(1, 1)} & -B_z\gamma^{(\mathrm{n})}_{1,1} & 0 \\
    \hat{F}_x^{(1, 2)} & \hat{F}_y^{(1, 2)} & \hat{F}_z^{(1, 2)} & \hat{B}^{(1, 2)} & 1
  \end{bmatrix}
\end{equation}
where
\begin{equation}
  \hat{F}_r^{(i,j)} = A_{rx}^{(i,j)}\hat{I}_x + A_{ry}^{(i,j)}\hat{I}_y + A_{rz}^{(i,j)}\hat{I}_z \; (r=x,y,z)
\end{equation}
and
\begin{equation}
  \hat{B}^{(i,j)} = -B_x\gamma^{(\mathrm{n})}_{i,j}\hat{I}_x - B_y\gamma^{(\mathrm{n})}_{i,j}\hat{I}_y - B_z\gamma^{(\mathrm{n})}_{i,j}\hat{I}_z.
\end{equation}
For example, $\hat{W}_2=\hat{W}\substack{\sigma_2^\prime\\\beta_1\beta_2\\\sigma_2}$ is a tensor with 4 different indices, and its element with index $(\beta_1,\beta_2,\sigma_2^\prime,\sigma_2)=(2,3,:,:)$ is $A_{zy}^{(1, 1)}\delta_{\sigma_2^\prime,\sigma_2}$ and $(\beta_1,\beta_2,\sigma_2^\prime,\sigma_2)=(4,2,:,:)$ is $\hat{F}_y^{(1, 2)}$.
The third core tensor, corresponding to the $(i,j)=(1,3)$ nucleus, is given by
\begin{equation}
  \hat{W}_3 =
  \begin{bmatrix}
    1 & 0 & 0 & 0 & 0 \\
    0 & 1 & 0 & 0 & 0 \\
    0 & 0 & 1 & 0 & 0 \\
    0 & 0 & 0 & 1 & 0 \\
    \hat{F}_x^{(1, 3)} & \hat{F}_y^{(1, 3)} & \hat{F}_z^{(1, 3)} & \hat{B}^{(1, 3)} & 1
  \end{bmatrix}
\end{equation}
The fourth core tensor, corresponding to the electronic spins, is given by
\begin{equation}
  \hat{W}_4 =
  \begin{bmatrix}
    0 & 0 & 0 & 0 & \hat{S}_x^{(1)}\\
    0 & 0 & 0 & 0 & \hat{S}_y^{(1)}\\
    0 & 0 & 0 & 0 & \hat{S}_z^{(1)}\\
    0 & 0 & 0 & 0 & 1\\
    \hat{S}_x^{(2)} & \hat{S}_y^{(2)} & \hat{S}_z^{(2)} & 1 & \hat{H}_{\mathrm{ele}}
  \end{bmatrix}
\end{equation}
where $\hat{H}_{\mathrm{ele}}$ is the $4\times 4$ Hamiltonian of the electronic spins, including the Zeeman term, exchange term, dipolar term, and Haberkorn relaxation term.
The fifth core tensor, corresponding to the $(i,j)=(2,1)$ nucleus, is given by
\begin{equation}
  \hat{W}_5 =
  \begin{bmatrix}
    1 & 0 & 0 & 0 & \hat{F}_x^{(2, 1)}\\
    0 & 1 & 0 & 0 & \hat{F}_y^{(2, 1)}\\
    0 & 0 & 1 & 0 & \hat{F}_z^{(2, 1)}\\
    0 & 0 & 0 & 1 & \hat{B}^{(2, 1)}\\
    0 & 0 & 0 & 0 & 1
  \end{bmatrix}
\end{equation}
The sixth core tensor, corresponding to the $(i,j)=(2,2)$ nucleus, is given by
\begin{equation}
  \hat{W}_6 =
  \begin{bmatrix}
    A_{xx}^{(2, 3)} & A_{yx}^{(2, 3)} & A_{zx}^{(2, 3)} & \hat{F}_x^{(2, 2)} \\
    A_{xy}^{(2, 3)} & A_{yy}^{(2, 3)} & A_{zy}^{(2, 3)} & \hat{F}_y^{(2, 3)} \\
    A_{xz}^{(2, 3)} & A_{yz}^{(2, 3)} & A_{zz}^{(2, 3)} & \hat{F}_z^{(2, 2)} \\
    -B_x \gamma_{2, 3}^{(n)} & -B_y \gamma_{2, 3}^{(n)} & -B_z \gamma_{2, 3}^{(n)} & \hat{B}^{(2, 2)} \\
    0 & 0 & 0 & 1
  \end{bmatrix}
\end{equation}
and finally, the seventh core tensor, corresponding to the $(i,j)=(2,3)$ nucleus, is given by
\begin{equation}
  \hat{W}_7 =
  \begin{bmatrix}
    \hat{I}_x \\
    \hat{I}_y \\
    \hat{I}_z \\
    1
  \end{bmatrix}.
\end{equation}
Extension to arbitrary nuclear counts follows by repeating the patterns of $\hat{W}_3$ and $\hat{W}_5$.
Expanding $W_1$ to $W_7$ recovers the total Hamiltonian. For vMPDO/LPMPS, $\hat{H} \otimes \hat{\mathbb{1}}$ is encoded analogously.
We note that inappropriate energy units can cause numerical instability because MPOs mix identity blocks with physical coefficients (e.g., hyperfine tensors).
For more general models (e.g., nucleus-nucleus couplings), the same graph-matching scheme applies \cite{renGeneralAutomaticMethod2020a}. We provide an implementation at \url{https://github.com/KenHino/PyMPO}.
\subsection*{Lindblad master equation with tensor networks}
\label{subsec:lindblad-tn}
When Lindblad jumps are present in Section \ref{subsec:radicalhop}, we propagate using an Arnoldi integrator for the non-Hermitian Liouvillian. We also examined a split-operator variant
\begin{equation}
\begin{aligned}
\label{eq:decomposeL}
    \exp\left(\mathcal{P}_{T_{\rho(t)}\mathcal{M}}\left(\dhat{\mathcal{L}}-\dhat{L}\right)\frac{\Delta t}{2}\right)
    \exp\left(\dhat{L}\Delta t\right)
    \\
    \times \exp\left(\mathcal{P}_{T_{\rho(t)}\mathcal{M}}\left(\dhat{\mathcal{L}}-\dhat{L}\right)\frac{\Delta t}{2}\right)
    + \mathcal{O}(\Delta t^3)
\end{aligned}
\end{equation}
where $\dhat{L}$ is the jump superoperator, and $\dhat{\mathcal{L}}$ is the full Liouvillian. This can be advantageous when
\begin{enumerate}
    \item  the residual part $\dhat{\mathcal{L}}-\dhat{L}$ is Hermitian, which keeps the Hessenberg matrix in the Krylov subspace tridiagonal and thus accelerates the computation
    \item the norm of the Lindblad jump superoperator $\dhat{L}$ is significantly larger than that of the residual $\dhat{\mathcal{L}}-\dhat{L}$, which would otherwise impede the convergence of the Krylov method; and
    \item $\dhat{L}$ acts only on a single site, in which case an exact expression for the propagator $\exp\!\left(\dhat{L}\Delta t\right)$ is available and involves applying a one-site gate to the MPDO.
\end{enumerate}
We have tried this approach and found that the additional Trotter error in Eq~(\ref{eq:decomposeL}) is not negligible because of the large magnitude of $\dhat{L}$.
Another approach to treat the Lindblad master equation with locally purified tensor network introduces a renormalisation of the ancilla dimension, called Kraus dimension $K_{\mathrm{Kraus}}$~\cite{wernerPositiveTensorNetwork2016}.
This approach decomposes the dissipator into a sum of products of Kraus operators $\{\hat{B}_q\}$,
$\exp\left(\dhat{L}\Delta t\right) = \sum_{q=1}^{k} \hat{B}_{q} \otimes \hat{B}_{q}^{\ast}$.
At each time step, dissipation is applied via the operators $\hat{B}_q$, followed by renormalisation of the Kraus dimension from $K_{\mathrm{Kraus}}\times k$ back to $K_{\mathrm{Kraus}}$.

From our benchmark calculations, we found that the bond dimension must be taken as large as possible; introducing additional renormalisation indices was therefore prohibitive, and it has the same additional Trotter error in Eq.~(\ref{eq:decomposeL}).
With a time step of $\Delta t=0.25\;\ns$, the Krylov subspace propagation converged reliably.
We therefore employed the straightforward vectorisation approach with an Arnoldi integrator.

\section{Energy diagram of radical pair system without nuclei}
\begin{figure}
    \centering
    \includegraphics[width=0.8\linewidth]{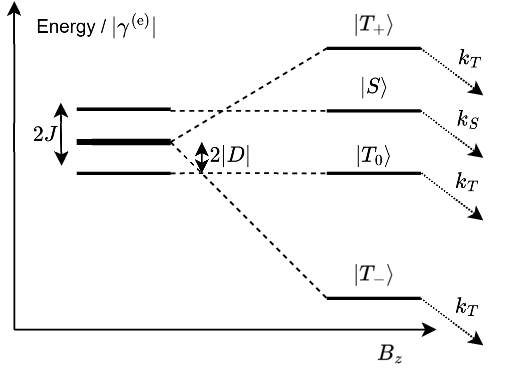}
    \caption{
    \label{fig:rp}
    Typical energy diagram of radical pair system}
\end{figure}

To familiarise the reader with the background of spin chemistry, we shall introduce the typical energy diagram of a radical pair system without nuclei.
Let $\ket{\Theta}$ be the two electronic spin states in the singlet-triplet basis.
Since they are eigenstates of $\hat{\mathbf{S}}_i^2$ and $\hat{\mathbf{S}}^2$, we have
\begin{equation}
  \hat{\mathbf{S}}_i^2 \ket{\Theta} = \frac{1}{2}\left(\frac{1}{2}+1\right)\ket{\Theta} = \frac{3}{4}\ket{\Theta}
\end{equation}
and
\begin{equation}
  \hat{\mathbf{S}}^2 \ket{\Theta} =
  \begin{cases}
    0(0+1)\ket{\Theta} = 0\ket{\Theta} & \text{if } \Theta = S \\
    1(1+1)\ket{\Theta} = 2\ket{\Theta} & \text{if } \Theta \in \{T_+, T_0, T_-\}
  \end{cases}
\end{equation}
Therefore, $\hat{\mathbf{S}}_1^\top\cdot\hat{\mathbf{S}}_2 = \frac{1}{2}\left(\hat{\mathbf{S}}^2-\hat{\mathbf{S}}_1^2-\hat{\mathbf{S}}_2^2\right)$ satisfies
\begin{equation}
\hat{\mathbf{S}}_1^\top\cdot\hat{\mathbf{S}}_2 \ket{\Theta} =
\begin{cases}
  -\frac{3}{4}\ket{\Theta} & \text{if } \Theta = S \\
  \frac{1}{4}\ket{\Theta} & \text{if } \Theta \in \{T_+, T_0, T_-\}
\end{cases}.
\end{equation}
Since the exchange term is defined as
$
  \hat{H}_{\mathrm{J}} = -J|\gamma^{(\mathrm{e})}| \left(
    2 \hat{\mathbf{S}}_1^\top \cdot \hat{\mathbf{S}}_2 - \frac{1}{2}\hat{\mathbb{1}}
  \right)
$
, we have
\begin{equation}
\frac{\braket{\Theta^\prime|\hat{H}_J|\Theta}}{|\gamma^{(\mathrm{e})}|} =
\begin{cases}
  2J & \text{if } \Theta^\prime = \Theta = S \\
  0 & \text{otherwise}
\end{cases}.
\end{equation}
Under the point-dipole approximation, the relative vector between two radicals is given by $\mathbf{r}=r\left(\hat{r}_x, \hat{r}_y, \hat{r}_z\right)^\top$, where $r=|\mathbf{r}|$, the dipolar term, is given by
\begin{equation}
\label{eq:dipolar}
  \begin{aligned}
  \hat{H}_{\mathrm{D}}
  &=
  \frac{\mu_0 \left(\gamma^{(\mathrm{e})}\right)^2}{4\pi}
  \left[\frac{\hat{\mathbf{S}}_1^\top\cdot\hat{\mathbf{S}}_2}{|\mathbf{r}|^3} -
  3 \frac{\left(\hat{\mathbf{S}}_1^\top\cdot \mathbf{r}\right)\left(\hat{\mathbf{S}}_2^\top \cdot \mathbf{r}\right)}{|\mathbf{r}|^5}
  \right]
    \\
  &= \frac{\mu_0 \left(\gamma^{(\mathrm{e})}\right)^2}{4\pi r^3}
  \hat{\mathbf{S}}_1^\top
  \cdot
  \begin{pmatrix}
  1 - 3\hat{r}_{x}^2 \;& -3\hat{r}_{x} \hat{r}_{y} \;& -3 \hat{r}_x\hat{r}_z \\
  - 3\hat{r}_{y}\hat{r}_x \;& 1 -3\hat{r}_{y}^2 \;& -3 \hat{r}_y\hat{r}_z \\
  - 3\hat{r}_{z}\hat{r}_x \;& -3\hat{r}_{z} \hat{r}_{y} \;& 1-3\hat{r}_z^2
  \end{pmatrix}
  \cdot
  \hat{\mathbf{S}}_2
  \\
  &=|\gamma^{(\mathrm{e})}| \hat{\mathbf{S}}_1^\top \cdot \mathbf{D} \cdot \hat{\mathbf{S}}_2
  \end{aligned}
\end{equation}
where
$\mathbf{D} \in \mathbb{R}^{3\times 3}$ denotes the dipolar coupling tensor between two electronic spins.
Particularly, by diagonalising $\mathbf{D}$, it can be characterised by a single scalar $D(r)$ and written as
\begin{equation}
\mathbf{D} = \frac{2}{3}D(r)\; \mathrm{diag}(-1, -1, 2)
\end{equation}
where
\begin{equation}
   D(r) = -\frac{3|\gamma^{(\mathrm{e})}|\mu_0}{8\pi r^3} \simeq -\frac{2786}{r^3} \, \mT \, \Ang^{-3}
\end{equation}
and $\mu_0$ is the magnetic permeability of vacuum \cite{santabarbaraBidirectionalElectronTransfer2005}.
As shown in Fig.~\ref{fig:rp}, the energy gap between singlet and triplet states is given by $2J$.
If the dipolar term under the point-dipole approximation is considered, the $T_+$ and $T_-$ states are lifted by $\tfrac{2}{3}|D|$, and the $T_0$ state is lowered by $\tfrac{4}{3}|D|$.
When a magnetic field is applied along the $z$-axis, the energy of $T_+$ is increased by $|\gamma^{(\mathrm{e})}|B_z$, and that of $T_-$ is decreased by $|\gamma^{(\mathrm{e})}|B_z$.
Since the projection operators $\hat{P}_S = \frac{1}{4}\hat{\mathbb{1}}_4 - \hat{\mathbf{S}}_1\hat{\mathbf{S}}_2$ and $\hat{P}_T = \hat{\mathbb{1}}_4 - \hat{P}_S$ satisfy
\begin{equation}
\begin{gathered}
  \hat{P}_S \ket{\Theta} =
  \begin{cases}
    \ket{\Theta} & \text{if } \Theta = S \\
    0 & \text{if } \Theta \in \{T_+, T_0, T_-\}
  \end{cases}
  \\
  \hat{P}_T \ket{\Theta} =
  \begin{cases}
    0 & \text{if } \Theta = S \\
    \ket{\Theta} & \text{if } \Theta \in \{T_+, T_0, T_-\}
  \end{cases},
\end{gathered}
\end{equation}
Haberkorn kinetics represent the kinetic pathways from singlet and triplet states with kinetic constants $k_S$ and $k_T$.
In particular, when $k_S = k_T = k$,
\begin{equation}
\frac{k_S}{2}\hat{P}_S + \frac{k_T}{2}\hat{P}_T = \frac{k}{2}\hat{\mathbb{1}}_4
\end{equation}
commutes with all other operators, which justifies the multiplication of an exponential decay factor $\exp\left(-kt\right)$ with the diagonal elements of the density matrix after propagation.
\section{Isotropic parameters for 18 nuclear spins}
In the benchmark simulation of the flavin anion and tryptophan cation radical pair system with isotropic hyperfine coupling presented in Section \ref{subsec:convergence}, we employed the isotropic hyperfine coupling constants listed in Table~\ref{tab:hyperfine}. These values are available in the \texttt{RadicalPy} library \cite{antillRadicalPyToolSpin2024}.

\begin{table}
  \centering
  \caption{Isotropic hyperfine coupling constants $a_{i,j}$ for the 18 nuclear spin system of flavin anion and tryptophan cation radical pair system.
  The value of tryptophan cation $i=2$ is identical to Ref~\cite{maedaMagneticallySensitiveLightinduced2012, hiscockQuantumNeedleAvian2016}.}
  \label{tab:hyperfine}
  \begin{ruledtabular}
  \begin{tabular}{ccccc}
    $i$ & $j$ & atom & $2I_{i,j}+1$ & $a_{i,j} / \mT$ \\
    \hline
    1  &  1  &  $\Hyd{1}$  & 2 & -0.1371 \\
    1  &  2  &  $\Hyd{1}$  & 2 & -0.1371 \\
    1  &  3  &  $\Hyd{1}$  & 2 & -0.1371 \\
    1  &  4  &  $\Nit{14}$ & 3 &  0.1784 \\
    1  &  5  &  $\Hyd{1}$  & 2 &  0.4233 \\
    1  &  6  &  $\Hyd{1}$  & 2 &  0.4263 \\
    1  &  7  &  $\Hyd{1}$  & 2 & -0.4403 \\
    1  &  8  &  $\Hyd{1}$  & 2 &  0.4546 \\
    1  &  9  &  $\Hyd{1}$  & 2 &  0.4546 \\
    1  &  10 &  $\Hyd{1}$  & 2 &  0.4546 \\
    1  &  11 &  $\Nit{14}$ & 3 &  0.5141 \\
    2  &  1  &  $\Hyd{1}$  & 2 &  1.605 \\
    2  &  2  &  $\Hyd{1}$  & 2 & -0.5983 \\
    2  &  3  &  $\Hyd{1}$  & 2 & -0.4879 \\
    2  &  4  &  $\Hyd{1}$  & 2 & -0.3634 \\
    2  &  5  &  $\Nit{14}$ & 3 &  0.3216 \\
    2  &  6  &  $\Hyd{1}$  & 2 & -0.278 \\
    2  &  7  &  $\Nit{14}$ & 3 &  0.1465 \\
  \end{tabular}
\end{ruledtabular}
\end{table}
\section{Size dependence of Monte Carlo ensembles}
\label{subsec:Kdeps}
\begin{figure}
  \centering
  \includegraphics[width=0.49\textwidth]{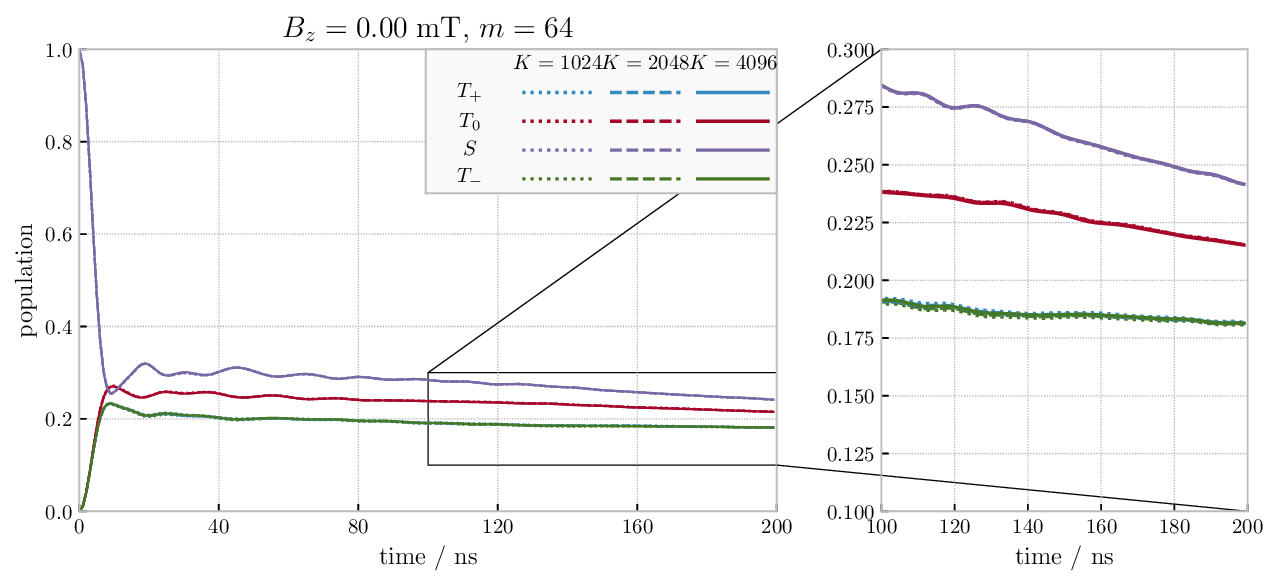}
  \includegraphics[width=0.49\textwidth]{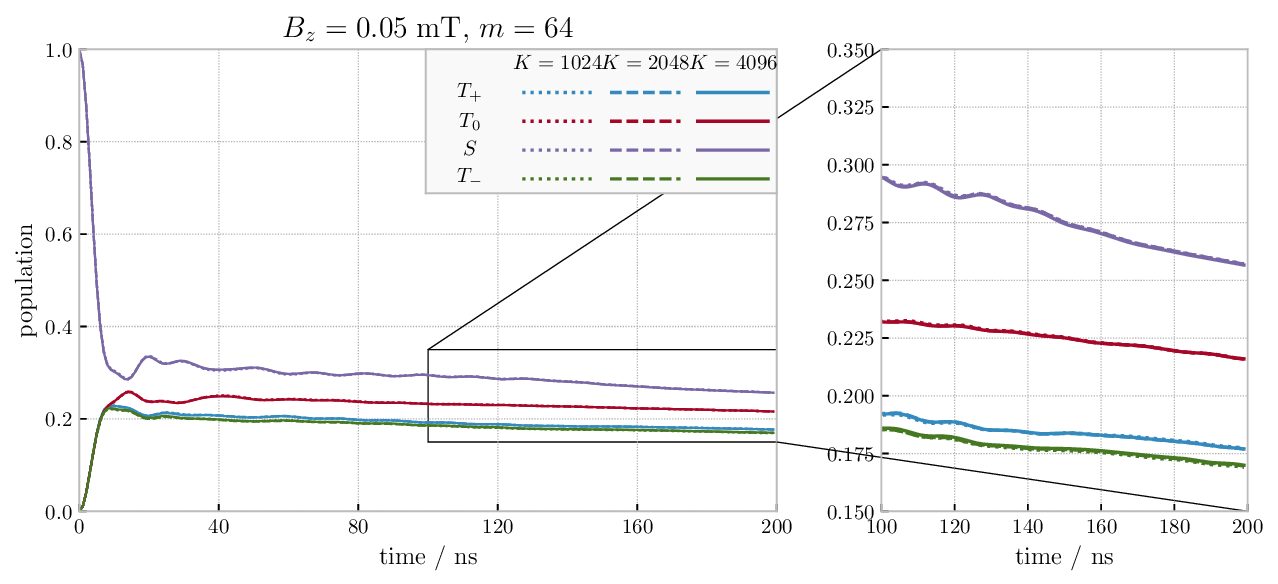}
  \includegraphics[width=0.49\textwidth]{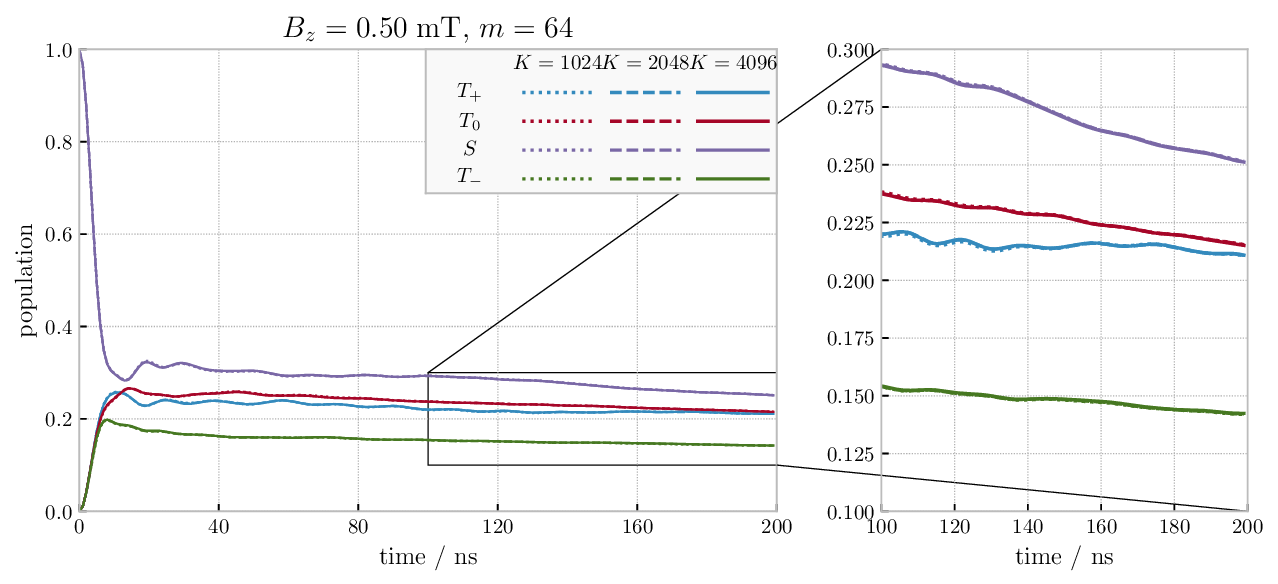}
  \includegraphics[width=0.49\textwidth]{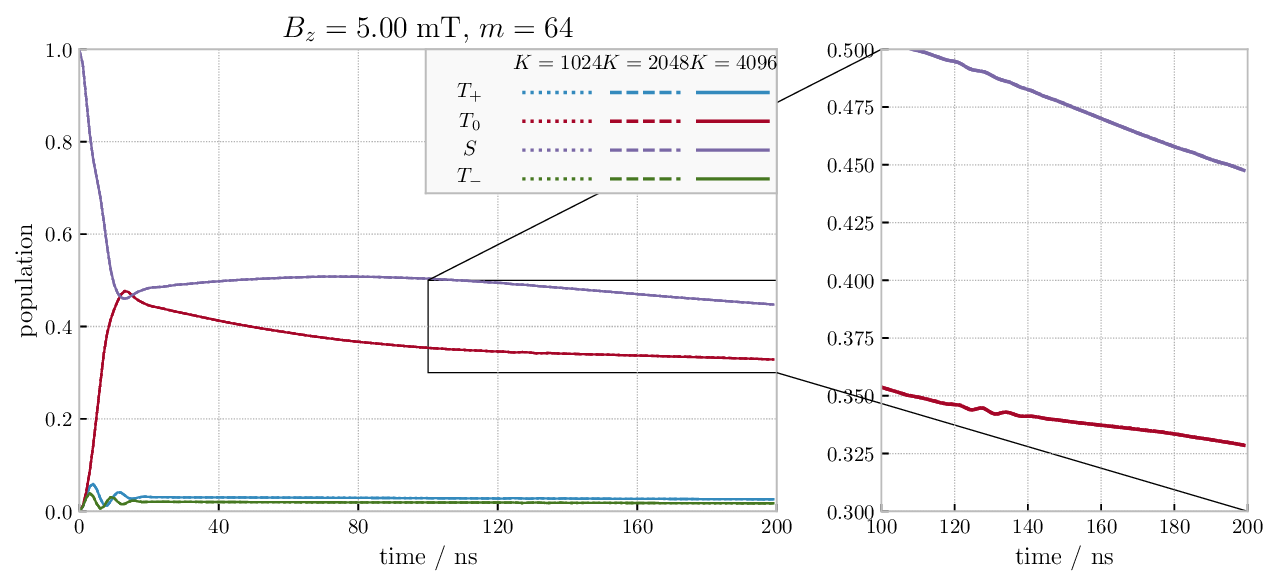}
  \includegraphics[width=0.49\textwidth]{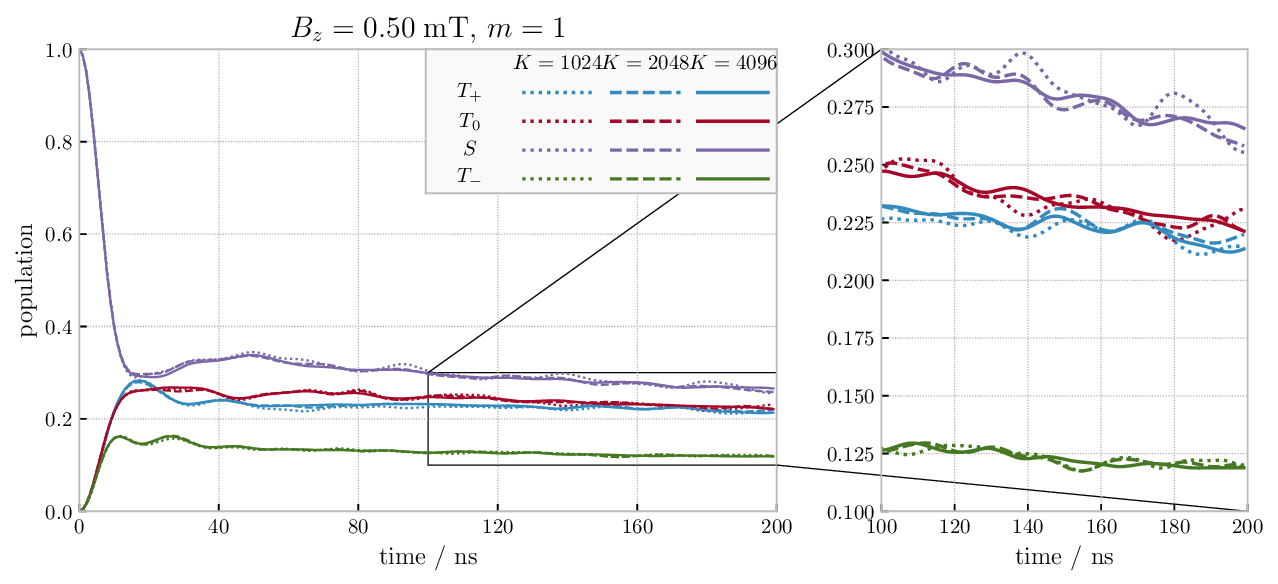}
  \caption{Convergence behaviour against the number of initial nuclear spin samples $K\in\{1024, 2048, 4096\}$ for $m=64$ (top 4 panels) and $m=1$ (bottom panel).
  }
  \label{fig:Kconv}
\end{figure}

For the stochastic full wavefunction method, we employed SU(Z) sampling \cite{faySpinRelaxationRadical2021,nemotoGeneralizedCoherentStates2000} to generate the initial nuclear spin state, while for the stochastic MPS method, we employed spin coherent state sampling \cite{radcliffePropertiesCoherentSpin1971}.
The spin-coherent state sampling approach employs
\begin{equation}
    \hat{\mathbb{1}} = \bigotimes_{j} \frac{2I_j+1}{4\pi} \int_0^{2\pi} \mathrm{d}\phi_j \int_0^\pi \mathrm{d}\theta_j \sin\theta_j
    \ket{\Omega_j}\bra{\Omega_j}
\end{equation}
where the initial $j$-th nuclear spin state is taken to be
\begin{equation}
    \ket{\Omega_j} = (1+|\zeta|^2)^{-I_j}e^{\zeta \hat{I}_{-}} \ket{I_j, M_I=+I_j}
\end{equation}
where $\zeta=e^{i\phi_j}\tan\left(\frac{\theta_j}{2}\right)$.
In general, SU(Z) sampling is more efficient than spin coherent state sampling; however, SU(Z) sampling assumes that one can access all configurations of the wavefunction, which is not feasible in tensor network methods. On the other hand, spin coherent state sampling requires only one-body spin sampling, which can be encoded into a rank-1 state of MPS.
We examined the number of initial samples $K$ needed to achieve sufficient convergence for the stochastic methods.
Fig.~\ref{fig:Kconv} shows the convergence behaviour with respect to the number of samples $K$ for $m=1$ and $m=64$.
We observed that $K=4096$ samples are sufficient to achieve convergence for $m=64$.
Interestingly, the computationally inexpensive mean-field treatment ($m=1$) is more sensitive to the initial nuclear spin configuration and thus requires more samples to achieve convergence.

\section{Equation of motion for the semiclassical approach}
\label{subsec:eom-sc}
The SC method demonstrated in Section \ref{subsec:convergence} is based on the second improved semiclassical theory by Fay et al.~\cite{p.fayHowQuantumRadical2020}.
The number of classical variables is $15+3N_1+3N_2$, where $N_1$ and $N_2$ are the numbers of nuclear spins coupled to the first and second electron spins, respectively.
First, we introduce two operators.
For any vector $\vec{v} = (v_x, v_y, v_z)^\top$, the operator $\wedge$ is defined by
\begin{equation}
    \vec{v}^{\wedge} = \begin{pmatrix}
    0 & -v_z & v_y \\
    v_z & 0 & -v_x \\
    -v_y & v_x & 0
    \end{pmatrix}
\end{equation}
and for any $3\times 3$ matrix $\mathbf{M}$, the operator $\vee$ is defined by
\begin{equation}
    \mathbf{M}^{\vee} = \begin{pmatrix}
      M_{yz} - M_{zy} \\
      M_{zx} - M_{xz} \\
      M_{xy} - M_{yx}
    \end{pmatrix}.
\end{equation}
We denote the interaction between two electrons by
\begin{equation}
  \mathbf{C} = |\gamma^{(\mathrm{e})}|\left(\mathbf{D} -2J\hat{\mathbb{1}}_3\right).
\end{equation}
From the Heisenberg equation of motion (EOM), $\frac{\mathrm{d}}{\mathrm{d}t} \hat{O}(t) = i\left[ \hat{H}_{\mathrm{total}}, \hat{O}(t) \right]$, we can derive the following EOM for the classical variables:
\begin{equation}
  \begin{aligned}
    \frac{\mathrm{d}}{\mathrm{d}t} \vec{S}_1 &= \vec{\omega}_1^{\mathrm{eff}} \times \vec{S}_1 + \left(\mathbf{C}\mathbf{T}^\top\right)^{\vee},\\
    \frac{\mathrm{d}}{\mathrm{d}t} \vec{S}_2 &= \vec{\omega}_2^{\mathrm{eff}} \times \vec{S}_2 + \left(\mathbf{C}^\top\mathbf{T}\right)^{\vee},\\
    \frac{\mathrm{d}}{\mathrm{d}t} \mathbf{T} &= \left(\vec{\omega}^{\mathrm{eff}}_1\right)^{\wedge} \mathbf{T}
    + \mathbf{T} \left\{\left(\vec{\omega}^{\mathrm{eff}}_2\right)^{\wedge}\right\}^{\top}
    \\
    &\quad- \frac{1}{4}\left[\left(\vec{S}_1\right)^{\wedge} \mathbf{C}+\mathbf{C}\left\{\left(\vec{S}_2\right)^{\wedge}\right\}^{\top}\right]\\
    \frac{\mathrm{d}}{\mathrm{d}t} \vec{I}_{i,j} &= \left(-\gamma_{i,j}^{\mathrm{(n)}}\vec{B} + |\gamma^{(\mathrm{e})}|\mathbf{A}_{i,j}\vec{S}_i\right) \times \vec{I}_{i,j}
  \end{aligned}
\end{equation}
where
\begin{equation}
  \vec{\omega}^{\mathrm{eff}}_i = -\gamma^{(\mathrm{e})} \vec{B} + \sum_{j=1}^{N_i} |\gamma^{(\mathrm{e})}| \mathbf{A}_{i,j} \vec{I}_{i,j}
\end{equation}
is the effective angular velocity for the $i$-th electron spin.
It is easier to interpret and implement in the following form:
\begin{equation}
  \begin{aligned}
\dot S_{1x} &= \omega_{1y} S_{1z} - \omega_{1z} S_{1y}
 + (C_{yx}T_{zx} + C_{yy}T_{zy} + C_{yz}T_{zz}) \\
 &\quad
 - (C_{zx}T_{yx} + C_{zy}T_{yy} + C_{zz}T_{yz}),\\
\dot S_{1y} &= \omega_{1z} S_{1x} - \omega_{1x} S_{1z}
 + (C_{zx}T_{xx} + C_{zy}T_{xy} + C_{zz}T_{xz}) \\
 &\quad
 - (C_{xx}T_{zx} + C_{xy}T_{zy} + C_{xz}T_{zz}),\\
\dot S_{1z} &= \omega_{1x} S_{1y} - \omega_{1y} S_{1x}
 + (C_{xx}T_{yx} + C_{xy}T_{yy} + C_{xz}T_{yz}) \\
 &\quad
 - (C_{yx}T_{xx} + C_{yy}T_{xy} + C_{yz}T_{xz}),\\
\dot S_{2x} &= \omega_{2y} S_{2z} - \omega_{2z} S_{2y}
 + (C_{xy}T_{xz} + C_{yy}T_{yz} + C_{zy}T_{zz}) \\
 &\quad
 - (C_{xz}T_{xy} + C_{yz}T_{yy} + C_{zz}T_{zy}),\\
\dot S_{2y} &= \omega_{2z} S_{2x} - \omega_{2x} S_{2z}
 + (C_{xz}T_{xx} + C_{yz}T_{yx} + C_{zz}T_{zx}) \\
 &\quad
 - (C_{xx}T_{xz} + C_{yx}T_{yz} + C_{zx}T_{zz}),\\
\dot S_{2z} &= \omega_{2x} S_{2y} - \omega_{2y} S_{2x}
 + (C_{xx}T_{xy} + C_{yx}T_{yy} + C_{zx}T_{zy}) \\
 &\quad
 - (C_{xy}T_{xx} + C_{yy}T_{yx} + C_{zy}T_{zx}),
\end{aligned}
\end{equation}
\begin{equation}
\begin{aligned}
 \dot T_{xx} &= \omega_{1y}T_{zx} - \omega_{1z}T_{yx}
           + \omega_{2y}T_{xz} - \omega_{2z}T_{xy} \\
           &\quad
           - \frac14(S_{1y}C_{zx} - S_{1z}C_{yx} + S_{2y}C_{xz} - S_{2z}C_{xy}),\\
\dot T_{yx} &= \omega_{1z}T_{xx} - \omega_{1x}T_{zx}
           + \omega_{2y}T_{yz} - \omega_{2z}T_{yy} \\
           &\quad
           - \frac14(S_{1z}C_{xx} - S_{1x}C_{zx} + S_{2y}C_{yz} - S_{2z}C_{yy}),\\
\dot T_{zx} &= \omega_{1x}T_{yx} - \omega_{1y}T_{xx}
           + \omega_{2y}T_{zz} - \omega_{2z}T_{zy} \\
           &\quad
           - \frac14(S_{1x}C_{yx} - S_{1y}C_{xx} + S_{2y}C_{zz} - S_{2z}C_{zy}),\\
\dot T_{xy} &= \omega_{1y}T_{zy} - \omega_{1z}T_{yy}
           + \omega_{2z}T_{xx} - \omega_{2x}T_{xz} \\
           &\quad
           - \frac14(S_{1y}C_{zy} - S_{1z}C_{yy} + S_{2z}C_{xx} - S_{2x}C_{xz}),\\
\dot T_{yy} &= \omega_{1z}T_{xy} - \omega_{1x}T_{zy}
           + \omega_{2z}T_{yx} - \omega_{2x}T_{yz} \\
           &\quad
           - \frac14(S_{1z}C_{xy} - S_{1x}C_{zy} + S_{2z}C_{yx} - S_{2x}C_{yz}),\\
\dot T_{zy} &= \omega_{1x}T_{yy} - \omega_{1y}T_{xy}
           + \omega_{2z}T_{zx} - \omega_{2x}T_{zz} \\
           &\quad
           - \frac14(S_{1x}C_{yy} - S_{1y}C_{xy} + S_{2z}C_{zx} - S_{2x}C_{zz}),\\
\dot T_{xz} &= \omega_{1y}T_{zz} - \omega_{1z}T_{yz}
           + \omega_{2x}T_{xy} - \omega_{2y}T_{xx} \\
           &\quad
           - \frac14(S_{1y}C_{zz} - S_{1z}C_{yz} + S_{2x}C_{xy} - S_{2y}C_{xx}),\\
\dot T_{yz} &= \omega_{1z}T_{xz} - \omega_{1x}T_{zz}
           + \omega_{2x}T_{yy} - \omega_{2y}T_{yx} \\
           &\quad
           - \frac14(S_{1z}C_{xz} - S_{1x}C_{zz} + S_{2x}C_{yy} - S_{2y}C_{yx}),\\
\dot T_{zz} &= \omega_{1x}T_{yz} - \omega_{1y}T_{xz}
           + \omega_{2x}T_{zy} - \omega_{2y}T_{zx} \\
           &\quad
           - \frac14(S_{1x}C_{yz} - S_{1y}C_{xz} + S_{2x}C_{zy} - S_{2y}C_{zx}).
  \end{aligned}
\end{equation}
The quantum projection operators onto $T_+, T_0, S, T_-$ are given by
\begin{equation}
  \begin{aligned}
  \hat{P}_{T_+} &= \frac{1}{4} + \frac{1}{2}\left(\hat{S}^1_z+\hat{S}^2_z\right) + \hat{S}^1_z\hat{S}^2_z, \\
  \hat{P}_{T_0} &= \frac{1}{4} - \hat{S}^1_x\hat{S}^2_x - \hat{S}^1_y\hat{S}^2_y + \hat{S}^1_z\hat{S}^2_z, \\
  \hat{P}_{S} &= \frac{1}{4} - \hat{S}^1_x\hat{S}^2_x - \hat{S}^1_y\hat{S}^2_y - \hat{S}^1_z\hat{S}^2_z, \\
  \hat{P}_{T_-} &= \frac{1}{4} - \frac{1}{2}\left(\hat{S}^1_z+\hat{S}^2_z\right) + \hat{S}^1_z\hat{S}^2_z.
  \end{aligned}
\end{equation}
By analogy with the above equations, the corresponding classical observables are evaluated by
\begin{equation}
  \begin{aligned}
    P_{T_+} &= \frac{1}{4} + \frac{1}{2}\left(S_{1z}+S_{2z}\right) + \mathbf{T}_{zz}, \\
    P_{T_0} &= \frac{1}{4} + \mathbf{T}_{xx} + \mathbf{T}_{yy} - \mathbf{T}_{zz}, \\
    P_{S} &= \frac{1}{4} - \mathbf{T}_{xx} - \mathbf{T}_{yy} - \mathbf{T}_{zz}, \\
    P_{T_-} &= \frac{1}{4} - \frac{1}{2}\left(S_{1z}+S_{2z}\right) + \mathbf{T}_{zz}.
  \end{aligned}
\end{equation}
The time evolution of the population of state $X$ from the singlet state is given by
\begin{equation}
\begin{aligned}
  &\Braket{P_{X}(t)} \\
  &= 4 \left(\frac{1}{4\pi}\right)^{N_1+N_2+2} \int \mathrm{d}\boldsymbol{\Omega} \exp\left(-kt\right)P_{X}(t; \boldsymbol{\Omega})P_{S}(0; \boldsymbol{\Omega})
\end{aligned}
\end{equation}
The integral over the initial angular momentum $\boldsymbol{\Omega}$ is evaluated using Monte Carlo integration from the three-dimensional sphere with a radius $\sqrt{S(S+1)}$ for the electronic spins and $\sqrt{I_{i,j}(I_{i,j}+1)}$ for the nuclear spins.
The initial correlation tensor is prepared as $\mathbf{T}(0)_{\alpha\beta} = S_1^\alpha(0) S_2^\beta(0)$.
This approach provides an exact solution when $k_S=k_T=k$ and hyperfine-coupled nuclei are absent.
References \cite{schultenSemiclassicalDescriptionElectron1978, manolopoulosImprovedSemiclassicalTheory2013, lewisAsymmetricRecombinationElectron2014, p.fayHowQuantumRadical2020} provide further details about SC theory. A Julia implementation is available at \url{https://github.com/KenHino/ElectronSpinDynamics.jl}.
\section{Scalability for number of nuclei}
\label{subsec:scalability}
\begin{figure}
    \centering
  \begin{overpic}[width=.98\linewidth]{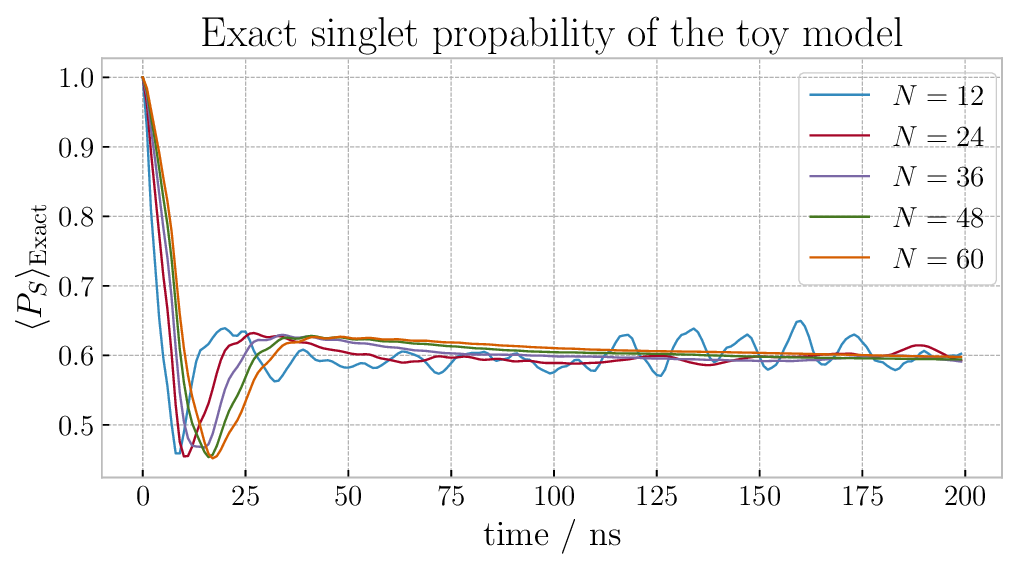}
    \put(1,50){\textbf{a}} % x=5 %, y=4 % from bottom-left
  \end{overpic}
  \begin{overpic}[width=.98\linewidth]{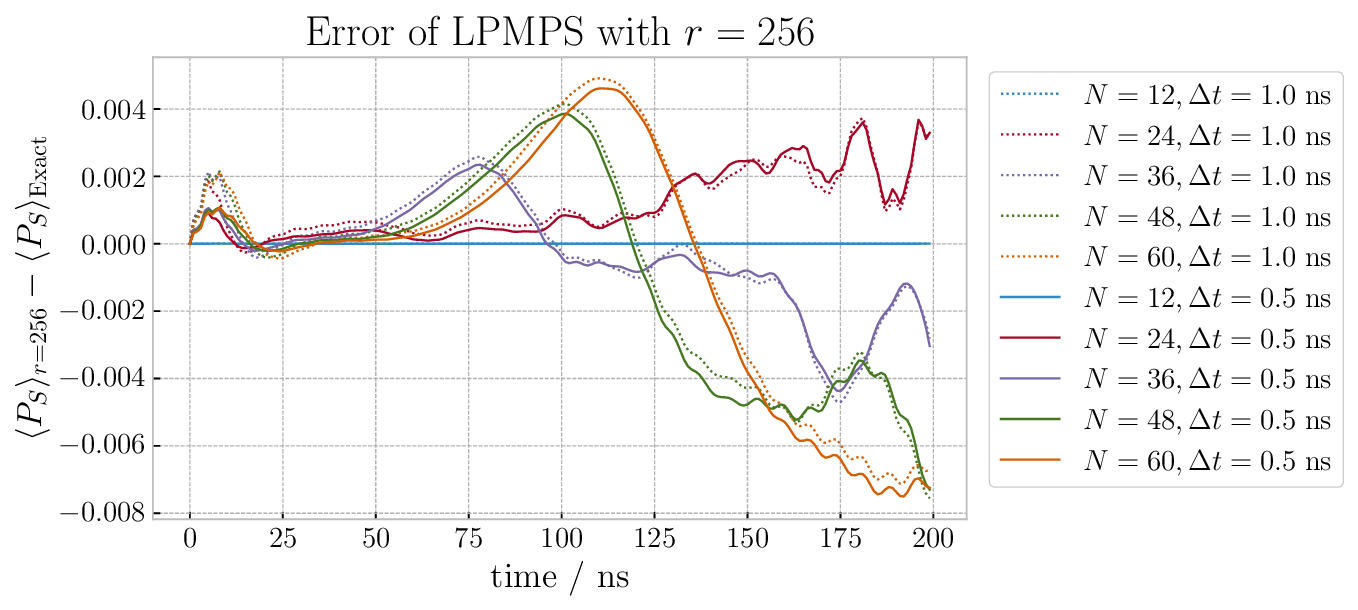}
    \put(1,45){\textbf{b}} % x=5 %, y=4 % from bottom-left
  \end{overpic}
    \begin{overpic}[width=.98\linewidth]{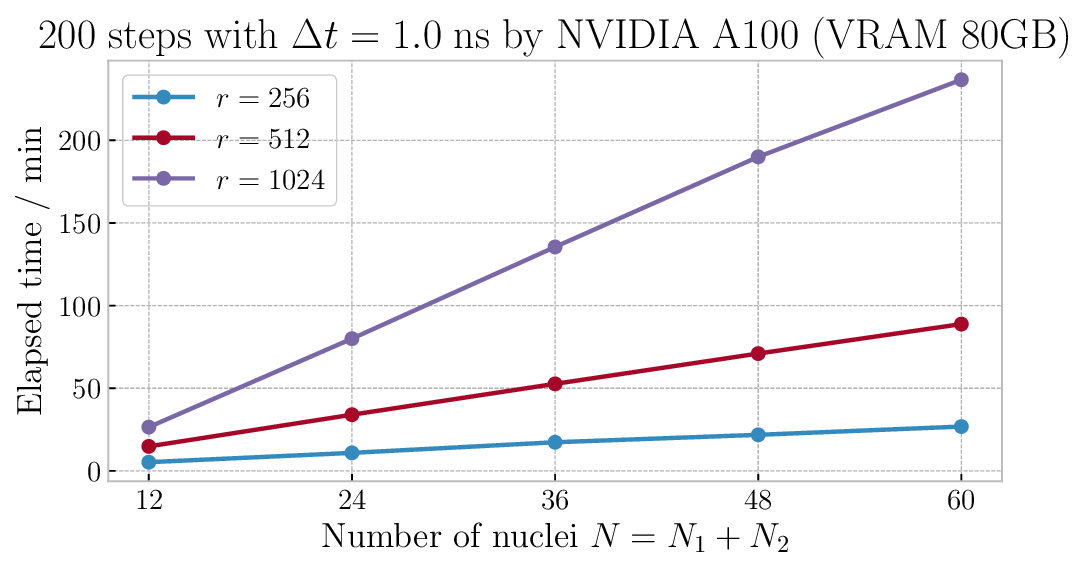}
    \put(1,50){\textbf{c}} % x=5 %, y=4 % from bottom-left
  \end{overpic}
    \caption{
    \textbf{(a)}: Singlet probability of numerically exact solution.
    \textbf{(b)}: Error between exact solution and LPMPS with $r=256$.
    \textbf{(c)}: Elapsed time for 200 ns propagation by LPMPS.
    }
    \label{fig:scale}
\end{figure}
As noted above, the actual wall-clock time is critical for practical simulations.

In this section, we demonstrate scalability concerning the number of nuclear spins.
We consider a toy radical pair comprising $N_1$ identical hydrogens in molecule $i=1$ and $N_2$ identical hydrogens in molecule $i=2$, for which the total Liouville-space dimension is $2^{2(N_1+N_2+2)}$.
To enable comparison with the exact solution, we set the isotropic hyperfine couplings to $a_{1j}=\tfrac{3.0}{N_1}\;\mT$ and $a_{2j}=\tfrac{9.0}{N_2}\;\mT$.
We scale the hyperfine tensors by the number of nuclei because, as the system size grows, the electron density is distributed over more sites, and the contribution of each nucleus decreases.
The exchange coupling is set to $J=2.5\;\mT$, the dipolar coupling is set to zero, and the magnetic-field strength is set to $|B|=5.0\;\mT$.
The LPMPS simulation uses $r=256$, a time step of $1\;\ns$, and a total propagation time of $200\;\ns$.
Exact simulations use symmetry-reduction techniques \cite{lindoySimpleAccurateMethod2018}, which decompose the problem into $\left(\left\lfloor\tfrac{N_1}{2}\right\rfloor+1\right)\!\times\!\left(\left\lfloor\tfrac{N_2}{2}\right\rfloor+1\right)$ independent sectors with a maximum Liouville-space dimension of $[4(N_1+1)(N_2+1)]^2$. The details are shown in Appendix \ref{subsec:symmetry-reduction}.
Fig.~\ref{fig:scale}a shows the exact singlet population dynamics in this toy model.
Clear oscillations are visible for small numbers of hyperfine-coupled nuclei, whereas an increasing number of nuclei yields a smoother signal.
Fig.~\ref{fig:scale}b shows the deviation of the singlet population by LPMPS with $r=256$ relative to the exact solution.
For $N=12$, which exhibits non-trivial population oscillations and, without low-rank approximation, would require a bond dimension of $4^6=4096$. We could reproduce the dynamics accurately using only a bond dimension $r=256$.
Furthermore, the absolute accuracy remains a 1\% deviation even as the system size increases.
The slight deviation in the beginning, around 10 ns, is due to the Trotter decomposition error, which a shorter time step size has mitigated.
Fig.~\ref{fig:scale}c shows the measured wall-clock time, which scales linearly with $N=N_1+N_2$ and cubically with $r$.
Our implementation uses the \texttt{JAX} library \cite{jax2018github}, enabling GPU acceleration. Computations were performed on an NVIDIA A100 GPU (VRAM 80 GB).
We note that to isolate scalability, just-in-time compilation time is excluded from the measurements by executing the initial time step in advance.
\section{Symmetry reduction for exact simulation}
\label{subsec:symmetry-reduction}
\begin{figure}
  \centering
  \includegraphics[width=0.98\linewidth]{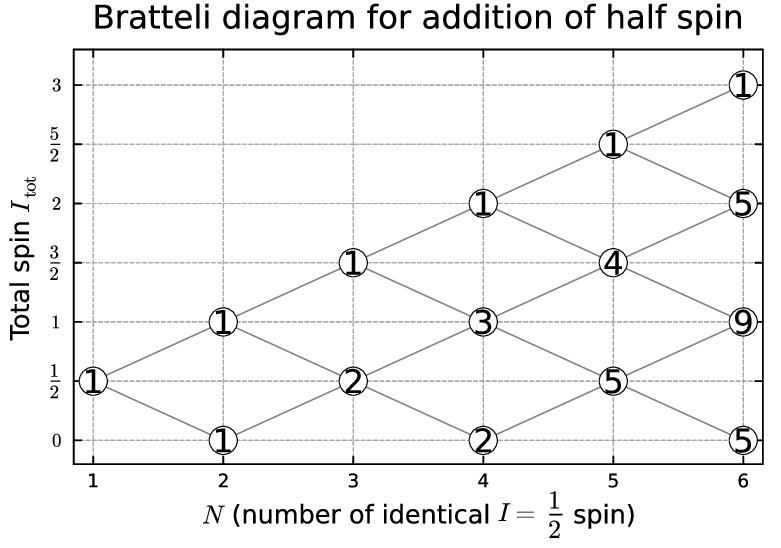}
  \caption{Bratteli diagram for the addition of half spin. The values in circles represent degree of degeneracy $D(N, I_{\mathrm{tot}})$.}
  \label{fig:bratelli}
\end{figure}
We employed the symmetry reduction technique
\cite{lindoySimpleAccurateMethod2018} when computing exact solutions in Appendix \ref{subsec:scalability}
and when effectively handling equivalent protons in methyl groups in Section \ref{subsec:radicalhop}.
When the system consists of $N$ identical nuclear spins $I_k\;(k=1,2,\cdots,N)$, which have the same gyromagnetic ratio $\gamma^{(\mathrm{n})}_k$ and hyperfine coupling tensor $\mathbf{A}_k$, the Zeeman term of the nuclear spins can be rewritten as
\begin{equation}
  \sum_{k=1}^{N} \gamma^{(\mathrm{n})}_k \mathbf{B} \cdot \hat{\mathbf{I}}_k
  = \gamma^{(\mathrm{n})} \mathbf{B} \cdot \hat{\mathbf{I}}_{\mathrm{tot}}
\end{equation}
where $\hat{\mathbf{I}}_{\mathrm{tot}} :=\sum_{k=1}^{N} \hat{\mathbf{I}}_k$ is the total nuclear spin operator.
The hyperfine term of nuclear spins can be rewritten as
\begin{equation}
  \sum_{k=1}^{N}
  \hat{\mathbf{S}}^\top \cdot \mathbf{A}_k \cdot \hat{\mathbf{I}}_k
  =
  \hat{\mathbf{S}}^\top \cdot \mathbf{A} \cdot \hat{\mathbf{I}}_{\mathrm{tot}}
\end{equation}
where $\hat{\mathbf{S}}$ is the electron spin operator coupled to the $k$-th nuclear spin with the hyperfine coupling tensor $\mathbf{A}_k$.

Since $\hat{\mathbf{I}}_{\mathrm{tot}}$ commutes with the total Hamiltonian and observables, the total nuclear spin $I_{\mathrm{tot}}$ is a ``good'' quantum number, with different values of $I_{\mathrm{tot}}$ being independent, and degenerate $I_{\mathrm{tot}}$ states being equivalent.
Fig.~\ref{fig:bratelli} shows the Bratteli diagram for the addition of half spin, which describes the degeneracy of the total nuclear spin $I_{\mathrm{tot}}$.
The accumulated weight of each sector is given by $D(N, I_{\mathrm{tot}})\frac{\left(2I_{\mathrm{tot}}+1\right)}{2^N}$, where $D(N, I_{\mathrm{tot}})$ is the degeneracy of the total nuclear spin $I_{\mathrm{tot}}$ for $N$ identical nuclear spins, which can be calculated using a recursion relation in a Bratteli graph or, when $I=\tfrac{1}{2}$, simply by
\begin{equation}
  D(N, I_{\mathrm{tot}}) = \binom{N}{\frac{N}{2}-I_{\mathrm{tot}}} - \binom{N}{\frac{N}{2}-I_{\mathrm{tot}}-1}.
\end{equation}
For example, when the system includes one methyl group, there are three equivalent $I=\tfrac{1}{2}$ proton spins. The naive treatment requires a nuclear spin Hilbert space dimension of $2^3=8$, and its Liouville space dimension is $8^2=64$.
By using symmetry reduction, the problem is decomposed into three independent sectors with $(I_{\mathrm{tot}}=\tfrac{3}{2})\oplus(I_{\mathrm{tot}}=\tfrac{1}{2})\oplus(I_{\mathrm{tot}}=\tfrac{1}{2})$, in which degenerate $I_{\mathrm{tot}}=\tfrac{1}{2}$ states are equivalent.
Therefore, one can evaluate the exact solution by solving 2 independent sectors with nuclear spin Hilbert space dimensions of $4$ and $2$.
By accumulating the results,
\begin{equation}
\begin{aligned}
  \Braket{P_X(t)} &= \frac{2\times \frac{3}{2}+1}{2^3}\Braket{P_X\left(t; I_{\mathrm{tot}}=\frac{3}{2}\right)} \\
  &\quad+ 2\times\frac{2\times \frac{1}{2}+1}{2^3}\Braket{P_X\left(t; I_{\mathrm{tot}}=\frac{1}{2}\right)},
\end{aligned}
\end{equation}
one can recover the exact solution.
As another example, $N_1=N_2=6$ in Appendix \ref{subsec:scalability}, the naive treatment requires a Hilbert space dimension of $4\times 2^{12}=16,384$, and its Liouville space dimension is $(16,384)^2=268,435,456$.

Whilst the tensor network method intentionally employs this naive treatment to demonstrate its efficiency, the exact solution is evaluated by accumulating the results of $4\times 4=16$ independent sectors with a Liouville space dimension of at most $\left[4\times (2\times3+1) \times (2\times3+1)\right]^2=196^2=38,416$.
\section{Anisotropic parameters for cryptochrome}
\label{subsec:aniso-params}
Exchange couplings are employed from the out-of-phase electron spin echo envelope modulation (ESEEM) measurement \cite{gravellSpectroscopicCharacterizationRadical2025},
\begin{equation}
    J_{12} = 0.011 \;\mT, J_{13} = 0.001\;\mT.
\end{equation}
To evaluate dipolar couplings by Eq~(\ref{eq:dipolar}),
we employed the relative position vector from the centre of mass (COM) of the aromatic ring of flavin to that of \Trp{C} and \Trp{D}.
From the crystal structure (PDB: 6PU0) \cite{zoltowskiChemicalStructuralAnalysis2019}, we employed
\begin{equation}
    \begin{gathered}
    \mathbf{r}_{12}=[ 9.480,\; -13.675,\; 5.388]^\top \;\mathrm{\AA},
    \\
    \mathbf{r}_{13}=[ 8.980,\; -18.684,\; 4.159]^\top \;\mathrm{\AA}.
    \end{gathered}
\end{equation}
Resulting electronic spin couplings are
\begin{equation}
    \begin{aligned}
    \mathbf{D}_{12}-2J_{12}\mathbb{1} &=
    \begin{pmatrix}
    -0.019 & -0.441 & -0.174 \\
    -0.441 &  -0.311 & 0.251 \\
     -0.174 & 0.251 & -0.226
    \end{pmatrix}
    \;\mT,
    \\
    \mathbf{D}_{13}-2J_{13}\mathbb{1} &=
    \begin{pmatrix}
    0.068 & 0.221 &  -0.049 \\
    0.221 & -0.286 & 0.102 \\
    -0.049 & 0.102 & 0.152
    \end{pmatrix}
    \;\mT,
    \end{aligned}
\end{equation}
which are of the same order of magnitude as the coupling with the nuclear spin bath.
We note that the masses of carbon and nitrogen were assumed to be equal when calculating the centres of mass (COMs).
These coordinates are given in the laboratory frame, which coincides with both the spin dynamics frame and the FAD frame shown in Fig.~\ref{fig:crypto}a.

To obtain the hyperfine couplings between electrons and nuclei, we performed electronic structure calculations using ORCA 6.0 \cite{neeseORCAProgramSystem2012}.
First, we extracted the coordinates of FAD, the W318 residue, and the W369 residue from the crystal structure of cryptochrome.
Then, we optimised the geometries of the flavin anion and tryptophan cation at the UKS $\omega$B97X-D4 / def2-TZVPD level, with constraints applied to the side-chain dihedral angles to match those of the crystal structure.
The optimised geometries were subsequently translated and rotated to minimise the root-mean-square deviation of the aromatic ring positions relative to the crystal structure.

Finally, we computed the hyperfine couplings at the UKS $\omega$B97X-D4 / EPR-III level.
The resulting hyperfine tensors include the Fermi contact term, dipolar coupling, and orbital contributions.
The calculated hyperfine tensors are listed in tables Tab.~\ref{tab:aiso-hyperfine-flavin} and Tab.~\ref{tab:aiso-hyperfine-TrpCD}.
The ORCA input files used for computing the anisotropic hyperfine coupling tensors for the flavin anion and tryptophan cations are shown below.
The resulting anisotropic hyperfine coupling tensors are shown in Table~\ref{tab:aiso-hyperfine-flavin} and \ref{tab:aiso-hyperfine-TrpCD}.
\begin{figure}[H]
    \centering
    \includegraphics[width=0.98\linewidth]{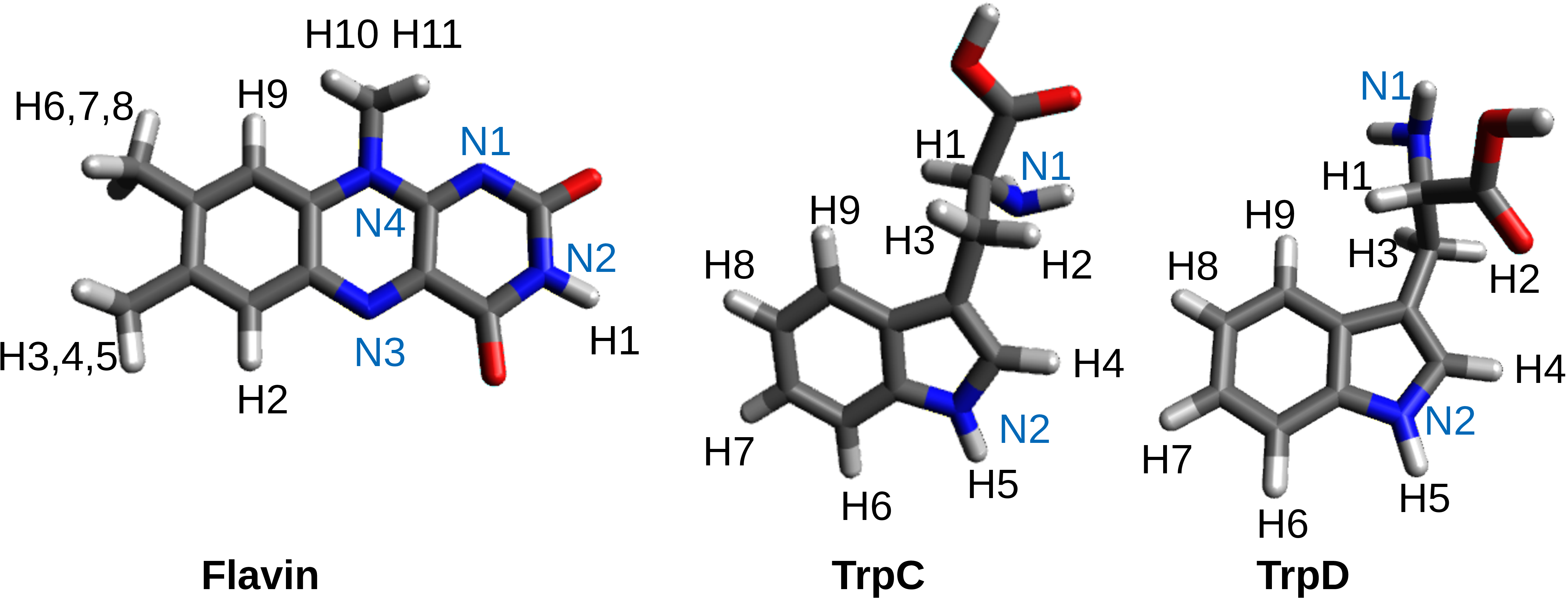}
    \caption{
    The label of hyperfine coupled nuclei corresponding to Tab.~\ref{tab:aiso-hyperfine-flavin} and \ref{tab:aiso-hyperfine-TrpCD}.
    The hydrogens that supposed to be side chains are omitted.
    }
    \label{fig:chromoCD}
\end{figure}
\begin{table}[H]
  \centering
  \caption{
  Anisotropic hyperfine coupling tensor of flavin anion. Hydrogens in the same methyl group are averaged over and regarded as isotropic, which are decomposed into the direct sum of $J=4$ and $J=2$ spins.
  %Side chain hydrogens are omitted. 
  See also Fig~\ref{fig:chromoCD} for label definition.
  }
  \label{tab:aiso-hyperfine-flavin}
  \begin{ruledtabular}
  \begin{tabular}{cc}
    label & flavin anion $A_{1,j} / \mT$ \\
    \hline
    N1 & $\begin{pmatrix*}[r] -0.022 & -0.002 & -0.003 \\ -0.002 & -0.019 & -0.005 \\ -0.003 & -0.005 &  0.043\end{pmatrix*}$ \\
    N2 & $\begin{pmatrix*}[r] -0.041 &  0.000 & -0.002 \\  0.000 & -0.030 & -0.000 \\ -0.002 & -0.000 & -0.066\end{pmatrix*}$ \\
    N3 & $\begin{pmatrix*}[r] -0.153 &  0.005 &  0.005 \\  0.005 & -0.143 &  0.034 \\  0.005 &  0.033 &  1.927\end{pmatrix*}$ \\
    N4 & $\begin{pmatrix*}[r] -0.002 &  0.002 &  0.079 \\  0.002 & -0.021 & -0.006 \\  0.079 & -0.006 &  0.627\end{pmatrix*}$ \\
    H1 & $\begin{pmatrix*}[r] -0.012 & -0.005 & -0.001 \\ -0.005 &  0.044 &  0.000 \\ -0.001 &  0.000 & -0.052\end{pmatrix*}$ \\
    H2 & $\begin{pmatrix*}[r] -0.199 & -0.046 &  0.003 \\ -0.046 & -0.547 & -0.003 \\  0.003 & -0.003 & -0.448\end{pmatrix*}$ \\
    H3,H4,H5 & -0.197 $\times\mathbb{1}_3$ \\
    %$\begin{pmatrix*}[r] -0.197 &  0.000 &  0.000 \\  0.000 & -0.197 &  0.000 \\  0.000 &  0.000 & -0.197\end{pmatrix*}$ \\
    H6,H7,H8 & 0.441 $\times \mathbb{1}_3$ \\
    %$\begin{pmatrix*}[r]  0.441 &  0.000 &  0.000 \\  0.000 &  0.441 &  0.000 \\  0.000 &  0.000 &  0.441\end{pmatrix*}$ \\
    H9 & $\begin{pmatrix*}[r]  0.097 &  0.033 & -0.004 \\  0.033 &  0.227 &  0.002 \\ -0.004 &  0.002 &  0.052\end{pmatrix*}$ \\
    H10 & $\begin{pmatrix*}[r]  0.238 &  0.027 & -0.045 \\  0.027 &  0.133 & -0.010 \\ -0.045 & -0.010 &  0.131\end{pmatrix*}$ \\
    H11 & $\begin{pmatrix*}[r]  0.195 & -0.033 & -0.040 \\ -0.033 &  0.101 &  0.012 \\ -0.040 &  0.012 &  0.086\end{pmatrix*}$ \\
  \end{tabular}
  \end{ruledtabular}
\end{table}
\begin{table}[H]
  \centering
  \caption{
  Anisotropic hyperfine coupling tensor of tryptophan cation. Hydrogens in the same methyl group are averaged over and regarded as isotropic. Protein ribbon bonded hydrogens are omitted. See also Fig~\ref{fig:chromoCD} for label definition.
  }
  \label{tab:aiso-hyperfine-TrpCD}
  \begin{ruledtabular}
  \begin{tabular}{ccc}
    label & \Trp{C} $A_{2,j} / \mT$ & \Trp{D} $A_{3,j} / \mT$\\
    \hline
  N1 & $\begin{pmatrix*}[r]  0.237 & -0.045 & -0.019 \\ -0.045 &  0.228 &  0.015 \\ -0.019 &  0.015 &  0.199\end{pmatrix*}$ & $\begin{pmatrix*}[r]  0.011 & -0.006 &  0.004 \\ -0.006 &  0.014 & -0.006 \\  0.004 & -0.006 &  0.009\end{pmatrix*}$ \\
  N2 & $\begin{pmatrix*}[r] -0.127 & -0.090 &  0.046 \\ -0.090 &  0.274 & -0.262 \\  0.046 & -0.262 &  0.027\end{pmatrix*}$ & $\begin{pmatrix*}[r] -0.151 &  0.001 &  0.005 \\  0.001 &  0.467 &  0.038 \\  0.005 &  0.038 & -0.131\end{pmatrix*}$ \\
  H1  & $\begin{pmatrix*}[r] -0.096 & -0.020 &  0.016 \\ -0.020 &  0.004 &  0.047 \\  0.016 &  0.048 & -0.160\end{pmatrix*}$ & $\begin{pmatrix*}[r] -0.095 & -0.080 &  0.001 \\ -0.080 & -0.062 & -0.059 \\  0.001 & -0.059 & -0.107\end{pmatrix*}$ \\
  H2  & $\begin{pmatrix*}[r]  0.606 & -0.021 & -0.084 \\ -0.021 &  0.562 &  0.078 \\ -0.084 &  0.078 &  0.714\end{pmatrix*}$ & $\begin{pmatrix*}[r]  0.062 &  0.015 & -0.082 \\  0.015 &  0.016 & -0.014 \\ -0.082 & -0.014 &  0.267\end{pmatrix*}$ \\
  H3  & $\begin{pmatrix*}[r]  0.476 & -0.007 & -0.102 \\ -0.007 &  0.218 &  0.007 \\ -0.102 &  0.007 &  0.279\end{pmatrix*}$ & $\begin{pmatrix*}[r]  1.386 &  0.029 &  0.086 \\  0.029 &  1.298 &  0.038 \\  0.086 &  0.038 &  1.522\end{pmatrix*}$ \\
  H4  & $\begin{pmatrix*}[r] -0.531 &  0.263 &  0.347 \\  0.263 & -0.546 &  0.107 \\  0.347 &  0.108 & -0.622\end{pmatrix*}$ & $\begin{pmatrix*}[r] -0.074 & -0.008 & -0.100 \\ -0.008 & -0.664 &  0.025 \\ -0.100 &  0.025 & -0.919\end{pmatrix*}$ \\
  H5  & $\begin{pmatrix*}[r] -0.594 & -0.104 & -0.114 \\ -0.104 & -0.338 &  0.203 \\ -0.114 &  0.203 & -0.091\end{pmatrix*}$ & $\begin{pmatrix*}[r] -0.369 & -0.022 &  0.343 \\ -0.022 & -0.452 & -0.018 \\  0.343 & -0.018 & -0.224\end{pmatrix*}$ \\
  H6  & $\begin{pmatrix*}[r] -0.676 & -0.138 & -0.187 \\ -0.138 & -0.525 &  0.098 \\ -0.187 &  0.098 & -0.348\end{pmatrix*}$ & $\begin{pmatrix*}[r] -0.662 & -0.019 &  0.230 \\ -0.019 & -0.550 & -0.014 \\  0.230 & -0.014 & -0.310\end{pmatrix*}$ \\
  H7  & $\begin{pmatrix*}[r]  0.084 &  0.029 &  0.018 \\  0.029 &  0.040 &  0.062 \\  0.018 &  0.062 &  0.089\end{pmatrix*}$ & $\begin{pmatrix*}[r]  0.153 &  0.000 &  0.008 \\  0.000 &  0.002 & -0.005 \\  0.008 & -0.005 &  0.077\end{pmatrix*}$ \\
  H8  & $\begin{pmatrix*}[r]  0.306 & -0.008 & -0.064 \\ -0.008 &  0.144 &  0.015 \\ -0.064 &  0.015 &  0.177\end{pmatrix*}$ & $\begin{pmatrix*}[r]  0.170 &  0.003 & -0.048 \\  0.003 &  0.137 & -0.015 \\ -0.048 & -0.015 &  0.316\end{pmatrix*}$ \\
  H9  & $\begin{pmatrix*}[r] -0.166 &  0.082 &  0.028 \\  0.082 & -0.580 & -0.133 \\  0.028 & -0.133 & -0.716\end{pmatrix*}$ & $\begin{pmatrix*}[r] -0.486 &  0.028 & -0.321 \\  0.028 & -0.499 & -0.003 \\ -0.321 & -0.003 & -0.456\end{pmatrix*}$ \\
  \end{tabular}
  \end{ruledtabular}
\end{table}
\begin{Verbatim}[fontsize=\footnotesize, frame=single, breaklines=false]
!UKS wB97X-D4 EPR-III TightSCF

# Flavin anion in optimised geometry

*xyz -1 2
N       -1.44910929       2.45682696       0.04714441
C       -0.79939824       3.65018192       0.02793392
O       -1.36283143       4.74306657       0.04244211
N        0.59503862       3.62101175      -0.01096827
C        1.41615850       2.50143785      -0.03227457
O        2.63399094       2.63776196      -0.06591721
C        0.68052887       1.24629698      -0.01139431
N        1.37266579       0.08286413      -0.03005704
C        0.64491416      -1.05672066      -0.01002260
C        1.30451618      -2.30134507      -0.02874900
C        0.62954209      -3.50936860      -0.01103714
C        1.39476090      -4.80903163      -0.03217088
C       -0.77161632      -3.50680028       0.02702427
C       -1.54829900      -4.79823278       0.04681639
C       -1.44926042      -2.28630996       0.04635646
C       -0.77284280      -1.07119517       0.02916144
N       -1.44302631       0.15103230       0.04958281
C       -0.73132358       1.34293713       0.02812803
C       -2.88989218       0.16594701       0.08764209
H        1.05861141       4.51578033      -0.02412850
H        2.38926924      -2.27264303      -0.05839870
H        2.47014003      -4.62265048      -0.05893608
H        1.13610042      -5.41586440      -0.90687141
H        1.18017944      -5.41793913       0.85289207
H       -1.34251114      -5.40962859      -0.83917460
H       -2.62311197      -4.60511089       0.07539243
H       -1.29603905      -5.41123278       0.91960783
H       -2.53205444      -2.30103704       0.07570838
H       -3.25407385      -0.34627691       0.98501152
H       -3.21170452       1.20301368       0.09995985
H       -3.30053392      -0.34044177      -0.79306288
*

%EPRNMR
    NUCLEI    = ALL H {AISO, ADIP, AORB}
    NUCLEI    = ALL N {AISO, ADIP, AORB}
END
\end{Verbatim}

\begin{Verbatim}[fontsize=\footnotesize, frame=single, breaklines=false]
!UKS wB97X-D4 EPR-III TightSCF

# Tryptophan C cation in optimised geometry

*xyz 1 2
N       13.18525047     -15.30694139       6.34165182
C       13.05015155     -15.38285425       4.89670705
C       14.33076846     -15.83599790       4.20048996
O       15.40768514     -15.84907470       4.72786330
C       12.66027700     -13.98828379       4.34283110
C       11.46691846     -13.51393894       5.08586110
C       11.50899930     -12.73528182       6.28558176
C       10.11722092     -13.88701382       4.91607705
N       10.29376083     -12.62248730       6.79151036
C        9.37648692     -13.32631672       5.98102013
C        9.45255764     -14.65565904       3.94382434
C        8.02610342     -13.49550656       6.12369248
C        8.08613410     -14.83788090       4.07307342
C        7.38720965     -14.26874487       5.14053330
O       14.10502622     -16.17639423       2.92264021
H       14.95052504     -16.43071765       2.52069039
H       14.14514555     -15.06596884       6.57491517
H       13.01269935     -16.21005098       6.76598401
H       12.25433855     -16.08681777       4.64135266
H       12.45666524     -14.05340577       3.27227777
H       13.49537730     -13.29760838       4.48871478
H       12.37227680     -12.28683465       6.75652387
H       10.06116479     -12.12195716       7.63680029
H        9.99572077     -15.08845606       3.11127714
H        7.46655361     -13.06444100       6.94549207
H        7.54722632     -15.42436502       3.33919178
H        6.31728120     -14.42600282       5.21558992
*

%EPRNMR
    NUCLEI    = ALL H {AISO, ADIP, AORB}
    NUCLEI    = ALL N {AISO, ADIP, AORB}
END
\end{Verbatim}

\begin{Verbatim}[fontsize=\footnotesize, frame=single, breaklines=false]
!UKS wB97X-D4 EPR-III TightSCF

# Tryptophan D cation in optimised geometry

*xyz 1 2
N        9.13550788     -16.82025191      -0.68031967
C        9.67414938     -16.85833920       0.66044269
C       10.85174920     -15.92279520       0.88042528
O       11.70122216     -16.13247980       1.71226721
C       10.12738538     -18.29565365       1.02460621
C       10.08815075     -18.53783527       2.48684086
C       11.24418843     -18.60383532       3.32991432
C        8.97480162     -18.60015251       3.35485720
N       10.87417430     -18.71606473       4.59240272
C        9.46416254     -18.71728005       4.67522790
C        7.58604442     -18.58191279       3.13724799
C        8.64978419     -18.80755802       5.77171386
C        6.74668054     -18.67030421       4.23480183
C        7.26799740     -18.78041569       5.52626417
O       10.81722745     -14.83722943       0.10940490
H       11.57943551     -14.27774580       0.32614492
H        9.16465622     -15.89002451      -1.07641092
H        8.18241083     -17.15577166      -0.71296050
H        8.94410562     -16.53138797       1.42046663
H       11.11852314     -18.49004317       0.61626997
H        9.42898326     -18.97951435       0.53283155
H       12.28183192     -18.55287721       3.03186281
H       11.50505459     -18.77472963       5.37853898
H        7.18136970     -18.50876390       2.13431505
H        9.03621818     -18.89579942       6.78024765
H        5.67260902     -18.65884410       4.09606946
H        6.58749924     -18.84996534       6.36716644
*

%EPRNMR
    NUCLEI    = ALL H {AISO, ADIP, AORB}
    NUCLEI    = ALL N {AISO, ADIP, AORB}
END
\end{Verbatim}

%\nocite{*}
%\bibliography{oxford-spin}% Produces the bibliography via BibTeX.
\bibliography{ms}% Produces the bibliography via BibTeX.

%\bibliography{ms}% common bib file
% @software{jax2018github,
%   author = {James Bradbury and Roy Frostig and Peter Hawkins and Matthew James Johnson and Chris Leary and Dougal Maclaurin and George Necula and Adam Paszke and Jake Vander{P}las and Skye Wanderman-{M}ilne and Qiao Zhang},
%   title = {{JAX}: composable transformations of {P}ython+{N}um{P}y programs},
%   url = {http://github.com/jax-ml/jax},
%   version = {0.3.13},
%   year = {2018},
% }

\end{document}